\documentclass[aps,prl,reprint,superscriptaddress,longbibliography]{revtex4-1}

\usepackage{siunitx}
\usepackage{amsmath,bm}
\usepackage{amsbsy}
\usepackage{amssymb}
\usepackage{mathptmx, textcomp}
\usepackage{color}
\usepackage{braket}
\usepackage{graphicx}
\usepackage{textcomp}
\definecolor{bl}{rgb}{0, .1, .6}

\usepackage[colorlinks=true, citecolor = bl, linkcolor = bl, urlcolor=bl, pdfborder={0 0 0}]{hyperref}
\usepackage{notes2bib}
\begin{document}

\title{Collective shift in resonant light scattering by a one-dimensional atomic chain}

\author{Antoine Glicenstein}
\author{Giovanni Ferioli}
\author{Nikola \v Sibali\'c}
\author{Ludovic Brossard}
\author{Igor Ferrier-Barbut}\email{igor.ferrier-barbut@institutoptique.fr}
\author{Antoine Browaeys}
\affiliation{Universit\'e Paris-Saclay, Institut d'Optique Graduate School, CNRS, Laboratoire Charles Fabry, 91127, Palaiseau, France}
\begin{abstract}

We experimentally study resonant light scattering by a one-dimensional randomly filled chain of cold two-level atoms. By a local measurement of the light scattered along the chain, we observe constructive interferences in light-induced dipole-dipole interactions between the atoms. They lead to a shift of the collective resonance despite the average interatomic distance being larger than the wavelength of the light. This result demonstrates that strong collective effects can be enhanced by structuring the geometrical arrangement of the ensemble. We also explore the high intensity regime where atoms cannot be described classically. We compare our measurement to a mean-field, nonlinear coupled-dipole model accounting for the saturation of the response of a single atom.
\end{abstract}

\maketitle

Two scatterers illuminated by a resonant light field are coupled to each other as the field radiated by one acts on the other, giving rise to a light-induced resonant dipole-dipole interaction. In a disordered ensemble containing many emitters, the random relative phases of the radiated fields lead to destructive interferences suppressing the effect of interactions. Structuring the sample could allow recovering constructive interferences, thus enhancing  dipole interactions and shaping its collective coupling to resonant light \cite{Aba07,Bet16b,Sha17,Scu18,Bac18}. Cold atoms provide an interesting platform to study collective light-matter interaction, exhibiting negligible inhomogeneous broadening. Experiments on disordered samples of cold atoms already led to the observation  of collective effects in near-resonant light scattering \cite{Ben10,Pel14,Gue16,Ara16,Roo16,Jen16,Gue17,Brom16,Cor17,Jen18}. Realizing ordered atomic arrays to enhance the collective coupling to light requires controlled positioning of the individual atoms with sub-wavelength precision. This sets stringent experimental requirements, but provides new pathways to engineer strong collective light-matter coupling. For example, the interactions can lead to enhanced reflectivity for a single atomic layer \cite{Bet16b,Sha17,Fac18}, an effect recently demonstrated using ultra-cold atoms in two-dimensional optical lattices \cite{Rui20}. In 1D arrays, it was predicted that interactions induce sub-radiant transport in atomically thin wires~\cite{Plan15,Chu15,Nee19,Bet16a,Ase17}. These predictions rely on models based on linear coupled dipoles (e.g.~\cite{Jav14}), or small scale full quantum models~\cite{Ase17,Jon16,Wil20}. This restricts the analysis either to the weak driving limit where a classical model is valid, or to small ensembles of up to about a dozen atoms, where full quantum calculations can be done. Experimentally, collective scattering with one-dimensional systems has been observed with atoms trapped near nano-photonic waveguides or nano-fibers \cite{Yu14,Vet10,Sol17,Cor19,Pra19}, and with chains of up to 8 trapped ions~\cite{Mei14}.\par
In this work we study resonant light scattering by a one-dimensional chain of two-level atoms as considered theoretically, e.\,g.~in~\cite{Bet16a,Sut16,Kra16}. For this we present a platform realizing a {\it free-space}, 1D partially filled chain of up to 100 atoms. We measure the intensity spectrum of the light scattered perpendicular to it. By local and global measurements of the resonance frequency shift we show that collective constructive interferences in resonant dipole-dipole interactions lead to an enhancement of the shift with respect to random dense ensembles~\cite{Pel14,Jen16,Cor17,Jen18}. Finally, we extend our experiments beyond the weak driving limit, and observe a suppression of the interaction-induced shift. We compare our findings to a model based on nonlinear coupled dipoles \cite{doE20}.\par

\begin{figure}
\includegraphics{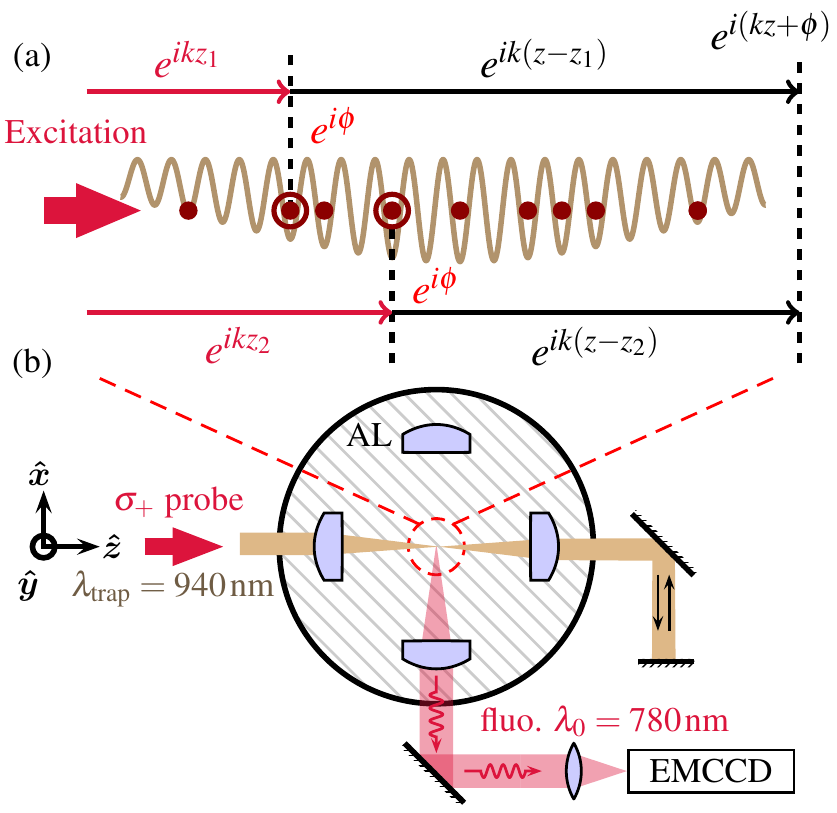} 
\caption{(a) Chain of atoms under axial excitation. 
The total phase accumulated by propagation and single scattering is the same in the forward direction irrespective of the position of the atom. This results in constructive interferences of all forward scattered fields. 
(b) Schematic of the experimental setup. Two orthogonal high-resolution optical systems based on 4 in-vacuum aspheric lenses (AL) realize a chain of single atoms in a 1D-optical lattice and collect the scattered light on an electron-multiplying CCD (EMCCD). \label{fig1}}
\end{figure}

To illustrate how the dimensionality of the atomic ensemble enhances collective scattering, consider a 1D chain of atoms excited by a plane wave (frequency $\omega=k\,c$) propagating along the chain axis $\boldsymbol{\hat{z}}$ (Fig.\,\ref{fig1}) as proposed in \cite{Sut16}. In the low-intensity limit, the dipoles respond linearly to the field $E$ \cite{Ruo97}, i.\,e.~the dipole  of atom $n$ at position $z_n$ is $d_n=\varepsilon_0\alpha E(z_n)$, with $\alpha=i(6\pi/ k_0^3)/(1-2i\Delta/\Gamma_0)$ the atomic polarizability. Here, $k_0=2\pi/\lambda_0$ is the transition wavevector, $\Delta=\omega-\omega_0$ the detuning with respect to the single-atom resonance frequency $\omega_0$, and $\Gamma_0$ the linewidth. By propagating along $\boldsymbol{\hat{z}}$, the driving field accumulates a phase $k\,z_n$ on atom $n$. The induced dipole $d_n$ scatters a field  phase-shifted by $\phi={\rm Arg}(\alpha)$ with respect to the driving plane wave~\cite{Alj09,Pot11,Cel11,Str12}. This scattered field accumulates a phase $k|z-z_n|$ by propagating along $z$. Therefore, in the forward direction ($z>z_n$), the phase accumulated by the field scattered by one atom is $k\,z+\phi$, independent of the atom's position. Now considering all atoms, the fields scattered in the forward direction are all in phase at first order (single scattering) and thus interfere constructively, as represented in Fig.\,\ref{fig1}(a). This conclusion only relies on the 1D geometry and holds even in the presence of position disorder along the chain. On the contrary, if the atoms are {\it not} aligned along the $\boldsymbol{\hat{z}}$ axis, the phases accumulated by the scattered fields do depend on their position, and their superposition in the forward direction does not lead to constructive interference.\par
To realize a 1D atomic chain and observe this effect, we introduce a new platform. We produce an optical lattice by retro-reflecting a tight optical tweezer focused by two in-vacuum aspheric lenses with numerical aperture NA =  0.5 [Fig.\,\ref{fig1}(a)]. It yields a chain of traps with small inter-trap spacing (trapping wavelength $\lambda_{\rm{trap}}=\SI{940}{\nano\meter}$ resulting in $\SI{470}{\nano\meter}$ spacing), similar to \cite{Alt03}, but with tight transverse confinement. The trap beam waist $w_{\rm{trap}} = \SI{3.3}{\micro\meter}$ (Rayleigh range $z_{\rm R} \simeq \SI{36}{\micro \meter}$) is chosen to avoid strong variations of the radial confinement along the chain while keeping small trap volume \cite{Sch01}. The dipole trap depth at the waist is $\sim 3$\,mK, corresponding to peak transverse and longitudinal oscillation frequencies of respectively $\omega_{\rho}=2\pi\times\SI{50}{\kilo\hertz}$ and $\omega_z=2\pi\times\SI{750}{\kilo\hertz}$. Another asset of our setup is the introduction of a second pair of aspheric lenses on a transverse axis as used in \cite{Bru19} for trapping and probing single atoms, allowing here for local measurements along the lattice axis. The resolution of this system is  $\sim \SI{1}{\micro\meter}$.\par
\begin{figure}
\includegraphics{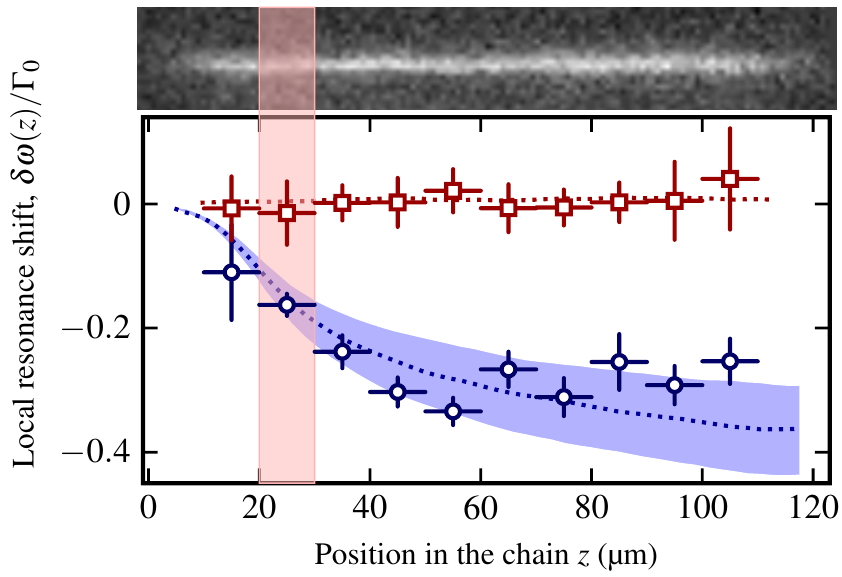} 
\caption{Local shift $\delta \omega(z)$ as a function of the position in the chain. Blue circles (red squares): axial (transverse) excitation. Each data point is the resonance frequency of a $\SI{10}{\micro\meter}$ segment around $z$. Horizontal error bars: segment width. Vertical error bars: standard error of the fit of the local spectrum. Dotted lines: results of coupled-dipoles simulations, with the shaded region corresponding to the experimental uncertainty in chain filling $\eta=0.5\pm 0.1$ .}\label{Fig2}
\end{figure}

We load the lattice with \textsuperscript{87}Rb atoms using the following sequence: We start from a 3D magneto-optical trap (MOT) superimposed to the lattice and then apply a $\SI{200}{\milli\second}$ $\Lambda$-enhanced grey molasses on the $D_1$ line \cite{Fer12,Gri13,Bro19} with the lattice tweezer on. We found empirically that applying the molasses results in a more reliable loading of the chain with respect to direct MOT loading. Thanks to the low photon scattering rate of grey molasses, the lattice is filled with an average of more than one atom per site. We then switch the MOT beams back on for 5\,ms. This pulse induces strong light-assisted collisions and ejects atoms out of shallow traps. The atoms are then optically pumped in the $\ket{5^2S_{1/2},F=2,m_F=2}$ state, with the quantization axis set by a 0.5 G magnetic field aligned with the chain. At the end of the loading sequence, the 200 central lattice sites are loaded with an average filling $\eta=0.5\pm 0.1$. The average interatomic distance is thus $\langle r_{\rm nn} \rangle\simeq\lambda_{\rm{trap}}\simeq1.2\,\lambda_0$ (here $\lambda_0\simeq\SI{780}{\nano\meter}$). The average loading is measured by illuminating the chain with a \emph{saturating} resonant beam in free flight and comparing the fluorescence of the whole chain (see example of an average image in Fig.\,\ref{Fig2}) with that of a single atom calibrated independently. The final temperature is $T\,=\SI{80(20)}{\micro \kelvin}$, yielding a transverse width $\sigma_{\rho}\simeq\SI{300}{\nano\meter}\simeq0.38\lambda_0$.

\begin{figure}
\includegraphics{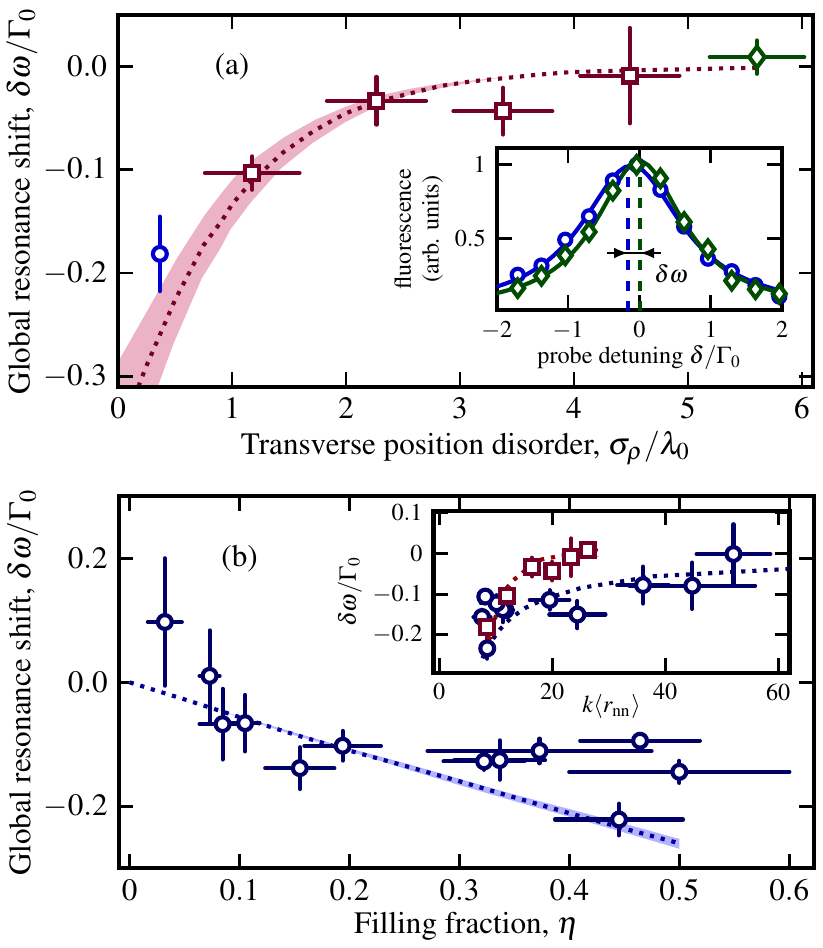} 
\caption{(a) Global shift $\delta\omega$ as a function of the radial size $\sigma_{\rho}$ of the cloud after a time-of-flight. Vertical error bars: fit standard errors, horizontal errors: size variation during probe pulse. Dotted line: coupled dipoles simulations accounting for the experimental uncertainty on the chain filling $\eta$. Inset: example of fluorescence spectra. (b) Global shift $\delta\omega$ vs.~$\eta$ compared to coupled dipole simulations (shaded region accounting for experimental uncertainty in temperature $\SI{80\pm 20}{\micro\kelvin}$. Vertical error bars: fit standard errors. Horizontal errors: experimental uncertainties. The reference of the shifts is the intercept of a linear fit of the data. Inset: comparing data of (a) and (b), plotted vs.~nearest-neighbour distance: $k\langle r_{\rm{nn}} \rangle$.}\label{fig3}
\end{figure}

Using this platform, we first explore collective scattering in the low-intensity limit. The atoms are excited along the chain axis by applying 200-ns pulses of a $\sigma_+$-polarized probe at $\lambda_{\rm{0}}=\SI{780.2}{\nano\meter}$ ($D_2$ line). The probe waist is $w_{\rm probe } = \SI{20}{\micro\meter}$ (Rayleigh range $z_{\rm{R}} = \SI{1.6}{\milli\meter}$) such that it approximates a plane wave. The probe intensity is $I/I_{\rm{sat}} \simeq0.3$. We image the light scattered by the atoms in the transverse direction on an electron multiplying CCD camera (EMCCD), through one of the two additional high-NA lenses. For a given probe detuning, the chain is illuminated by 50 probe pulses and we repeat  over 300 identically prepared samples to obtain sufficient statistics. The scattered intensity spectrum is extracted by repeating this at different detunings between $\Delta=-3\Gamma_0$ and $3\Gamma_0$. To reveal the effect of interactions along the chain, we divide it into $\SI{10}{\micro\meter}$-long segments, as shown on the top of Fig.\,\ref{Fig2}. We observe resonance profiles that are well fitted by a lorentzian, from which we extract the local shift of the resonance $\delta \omega(z)$ [inset Fig.\,\ref{fig3}(a)]. This {\it on-axis} excitation is compared to the result of an identical excitation procedure but with a plane wave probe ($w_{\rm probe \perp}\simeq\SI{1.5}{\milli\meter}$) sent {\it perpendicularly} to the chain. The results are plotted in Fig.\,\ref{Fig2}. Under perpendicular excitation, we do not observe any shift along the chain, while the shift does increase along the chain for the axial excitation, indicating a buildup of the interactions. These findings are in agreement with the qualitative discussion above. For comparison, shifts of comparable amplitude were obtained in disordered 2D and 3D samples but for interatomic distances about ten times smaller \cite{Pel14,Cor17}, highlighting the enhancement of the collective response by reducing the dimensionality. \par

We now describe our experimental results in terms of the the \emph{steady-state} coupled-dipole model \footnote{We have checked numerically that the shifts calculated by the steady-state model is the same as the one obtained by the time-dependent coupled-dipole model}. In this model, each atomic dipole of the chain is driven by the field of the plane wave and the sum of the fields radiated by all the other atoms: $d_n=\epsilon_0\alpha [E_{\rm L}({\boldsymbol r}_n)+ \sum_{m\neq n} G({\boldsymbol r}_n-{\boldsymbol r}_m) d_m]$ with $G(\boldsymbol{r})$ the Green's function \cite{SupMat}. Here, we assume scalar dipoles to reproduce the experimental arrangement of two-level atoms driven by a $\sigma_+$-polarized field \footnote{In 1D, the field of the  dipoles scattered along the chain are also $\sigma_+$-polarized, and therefore one does not need to resort to the application of a large magnetic field to isolate a two-level structure, as was done for a random ensemble~\cite{Jen18}. In this case the Green's function $G({\bf r})$ is also a scalar~\cite{SupMat}.}. To get an intuitive understanding of the shift increase along the chain, we first use a perturbative approach, as done in \cite{Sut16}. In the limit of large interparticle distance $(k_0\,\langle r_{\rm nn}\rangle>1)$, only the long-range part of the radiated field plays a role and $G(\boldsymbol r)\propto e^{ikr}/kr$. Keeping only forward scattering at first order (single scattering), the field intensity at position $z_n$ is \cite{Sut16,SupMat}:
\begin{equation}
|E^{\mbox{\tiny (1)}} (z_n,\Delta) | ^2 = | E_{\rm L} | ^2\left(1-\frac{6\Delta/\Gamma_0}{1+(2\Delta/\Gamma_0)^2}\sum_{z_m<z_n}\frac{1}{k|z_m-z_n|} \right)\ , \label{eqn:2}
\end{equation}
with $E_{\rm L}$ the laser field amplitude. This simple model shows that the field seen by atoms down the chain is increased for red detunings $(\Delta<0)$. Thus, the excitation probability $|d_n^{\mbox{\tiny (1)}}(\Delta)|^2\propto |E^{\mbox{\tiny (1)}}(z_n,\Delta) |^2/(1+(2\Delta/\Gamma_0)^2)$ is redshifted compared to the single-atom resonance due to the interactions. This interaction-induced shift is actually the equivalent of a collective Lamb shift \cite{Fri73} for a discrete medium \cite{Mei14,Rui20}. The interpretation is the following: for red detunings, the scattered and driving fields are in phase $(\phi < \pi/2)$ such that constructive interferences increase the field intensity $|E(z_n)|^2$\footnote{Note that although it may look like this argument implies guiding of the light along the chain, guided modes exist only for a spacing smaller than $\lambda_0/2$ see \cite{Bet16a,Ase17}.}. For blue detuning, they are out of phase $(\phi>\pi/2)$ and their destructive interferences reduce $|E(z_n)|^2$.\par

Though the simple perturbative model captures the mechanism behind the local shift, full solutions of the coupled dipoles including experimental imperfections are necessary for a quantitative comparison with the data. We thus numerically solve the set of {\it linear} coupled equations to calculate each dipole $d_n(\Delta)$ for various detunings. The power emitted by a dipole is proportional to ${\rm Im}[d_nE(z_n)^*]\propto|d_n|^2$. We take into account both the random filling fraction and the residual thermal fluctuation of the atomic positions (radially and axially) in each well by averaging over several hundreds of random realizations and plot the mean dipole in the chain slices used in the experiment, $\sum_{n\in {\rm slice}}|d_n(\Delta)|^2$, as a function of the detuning. The obtained spectra are well fitted by a Lorentzian lineshape, from which we extract the theoretical line-shift. The results are shown in Fig.\,\ref{Fig2}, for different fillings $\eta$ compatible with the experimental uncertainty. We obtain a good agreement between the data and the {\it ab-initio} model with no adjustable parameter.\par

Next, we vary the parameters controlling the interaction strength. As discussed above, the collective enhancement of interactions relies on the 1D geometry. Therefore, we first consider a situation away from 1D. In this case, if an atom is displaced by $\rho_n$ perpendicularly to the chain axis, the phase factor on axis is $k|\boldsymbol r-\boldsymbol r_n|\simeq k|z-z_n|+k\rho_n^2/(2|z-z_n|)$. The relevant factor for constructive interferences to occur at an axial distance $\Delta z$ should thus be that the Fresnel number $\sigma_{\rho}^2/\lambda_0\Delta z\ll1$, with $\sigma_{\rho}$ the radial extent. This shows that when $\sigma_{\rho}/\lambda_0\gg1$, interferences should disappear. To check this experimentally, we change the radial size of the atomic distribution by letting the chain expand in free flight. After this time-of-flight we send a near-resonant probe pulse for $\SI{10}{\micro\s}$ along the chain and collect again the light scattered in the transverse direction. We now record the scattered intensity summed over {\it all} the chain for various detunings and extract the global shift of the resonance frequency $\delta\omega$. Figure \ref{fig3}(a) shows the evolution of $\delta\omega$ as a function of $\sigma_{\rho}/\lambda_0$. As expected, the shift, and hence the interactions, vanishes when the atoms are not in a 1D geometry. The dotted lines correspond to coupled-dipoles simulations computed with our experimental parameters. They are in good agreement with the data.\par

In another set of experiments we increase the interatomic distance while keeping the 1D geometry, by reducing the filling fraction of the chain \cite{SupMat}. The global shift as a function of the filling of the chain is shown in Fig.\,\ref{fig3}(b), together with the coupled-dipoles simulations. We experimentally observe a reduction of the shift, as predicted. However, the calculated linear dependence is not clear in the data. This may be explained by a non-uniform filling along the chain. The same data are plotted in the insert as a function of the average interatomic nearest-neighbour distance $k\langle r_{\rm nn}\rangle$ and compared with the data of Fig.\,\ref{fig3}(a): at a given $k\langle r_{\rm nn}\rangle$, the shift is much stronger for a 1D sample. This again shows that collective scattering is enhanced in 1D.\par

\begin{figure}
\includegraphics[width=\linewidth]{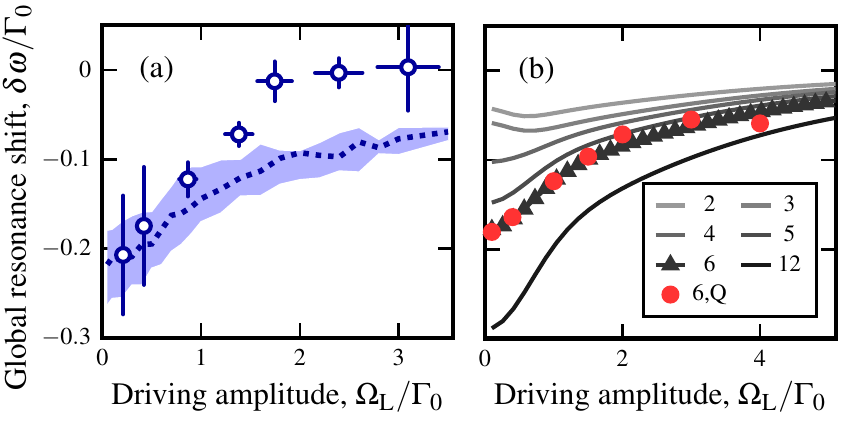} 
\caption{\label{fig4}
(a) Measured global resonance shift as a function of the laser Rabi frequency (circles). Vertical error bars from the fits. Horizontal errors: 10\% uncertainty on the probe intensity. Dotted lines: results of the NCD model including the experimental parameters. Shaded area: uncertainty in the filling fraction $\eta=0.5\pm0.1$. (b) Mean-field nonlinear coupled-dipole calculations for chains of $N$ atoms (solid lines) are in reasonable agreement with a full quantum model (circles).}
\end{figure}

Finally, we explore the evolution of the frequency shift when increasing the intensity of the driving field beyond the low-intensity limit. We again send the probe light along the chain axis and collect the transverse scattered light. We have verified that the higher intensity does not lead to significant extra atom losses and heating. We measure the scattered intensity spectrum integrated over the chain. When increasing the intensity of the probe light, we observe a broadening of the lorentzian line, as well as a suppression of the global shift as shown in Fig.\,\ref{fig4}(a).\par

To model the data, we use a nonlinear coupled-dipole (NCD) model accountings for the nonlinear single atom response (see also  \cite{doE20}): We solve again the coupled-dipole equations, but now using the nonlinear expression of the atomic polarizability for a strongly driven two-level atom given by the steady-state solution of the optical Bloch equations \cite{GAF}: $\alpha_{\rm NL}(\Delta,\Omega)=i\frac{(6\pi/ k_0^3)(1+2i\Delta/\Gamma_0)}{1+(2\Delta/\Gamma_0)^2+2\Omega^2/\Gamma_0^2}$. Here $\Omega = dE/\hbar$ ($d^2=3\pi\varepsilon_0\,\hbar\Gamma_0/k_0^3$) is the Rabi frequency and $E$ is the {\it total} field driving the atom, superposition of the laser field and the one scattered by all other atoms. This model amounts to a mean-field theory where the many-body density matrix is factorized into a product of individual atomic density matrices \cite{doE20,Kra15,Par18,SupMat}. To compare to the data, we calculate the fluorescence which is proportional to the sum of the excited state populations $\rho_{{\rm ee},n}={\rm Im}[\rho_{{\rm eg}, n}\Omega_n^*]/\Gamma_0$ of atoms in the chain (here $\rho_{{\rm eg},n}=d_n/2d$ is the atomic coherence between ground and excited states of atom $n$) \cite{GAF}. By solving this model, we observe that the spectrum becomes slightly asymmetric. To extract a shift nonetheless, we fit the center of the spectrum ($-2<\Delta/\Gamma_0<2$) with a lorentzian, as done in the experiments. Considering first unity-filled short chains, we obtain a reduction of the global shift with increased driving strength~[Fig.\,\ref{fig4}(b) solid lines]. To check the validity of this model, we compare it with the full solution that keeps quantum correlations into account~[Fig.\,\ref{fig4}(b) circles] calculated as in Refs.~\cite{Jon16,Ase17} for a chain of 6 atoms: the NCD model satisfyingly captures the evolution of the shift for the considered inter-atomic spacing.\par

The reduction of the resonance shift can be interpreted as an effect of the saturation of individual quantum emitters~\cite{Wil20}. Indeed, from the above nonlinear expression of the polarizability, the atomic dipole scales as $1/\Omega$ in the strong driving limit. Thus on a given atom, the ratio of the driving applied by the other ones to the external driving decreases as $1/\Omega^2$, hence suppressing interferences along the chain. The results of NCD calculations for the experimental parameters, involving $\sim 100$ of atoms, are shown in Fig.\,\ref{fig4}(a). They are in good agreement with the experimentally measured global shift for the weakest driving, but predict a more gradual tail-off to zero for stronger driving. Further investigations are required to elucidate the disagreement at large driving amplitudes. A further reduction of interferences could also be due to incoherent scattered light, and in the strong field limit some unaccounted-for mechanisms might depump atoms out of the two-level system.\par

The measurements presented here show that controlling the geometrical arrangement of an atomic sample allows to shape its collective response to light, as also observed in \cite{Rui20}. Further investigations beyond the classical regime of weak driving should follow. Increasing the coupling strength by reducing the interatomic distance will be a promising way forward to observe effects beyond what is captured by the mean-field model and to address long-lived subradiant states.  \par

\begin{acknowledgments}
We would like to thank, D. Barredo ,  F. Nogrette, Y. Sortais and W. Wu for assistance in early stages of the experiment, and T. Pohl, B. Olmos, J. Needham, J. Ruostekoski, D. Chang, R. Bettles,  A. Asenjo-Garcia and H. Ritsch for discussions. This work benefited from financial support by the R\'egion \^Ile-de-France in the framework of DIM Nano-K (project LISCOLEM) and DIM SIRTEQ (project DSHAPE), and by ``Investissements d'Avenir'' LabEx PALM (project ECONOMIQUE). N.\,\v{S}. is supported by EU Horizon 2020 (COQUDDE, Marie Sk{\l}odowska-Curie fellowship No 786702). A.\,G.~is supported by the D\'el\'egation G\'en\'erale de l'Armement, fellowship 2018.60.0027.
\end{acknowledgments}

\bibliography{chain}	

\begin{thebibliography}{55}%
\makeatletter
\providecommand \@ifxundefined [1]{%
 \@ifx{#1\undefined}
}%
\providecommand \@ifnum [1]{%
 \ifnum #1\expandafter \@firstoftwo
 \else \expandafter \@secondoftwo
 \fi
}%
\providecommand \@ifx [1]{%
 \ifx #1\expandafter \@firstoftwo
 \else \expandafter \@secondoftwo
 \fi
}%
\providecommand \natexlab [1]{#1}%
\providecommand \enquote  [1]{``#1''}%
\providecommand \bibnamefont  [1]{#1}%
\providecommand \bibfnamefont [1]{#1}%
\providecommand \citenamefont [1]{#1}%
\providecommand \href@noop [0]{\@secondoftwo}%
\providecommand \href [0]{\begingroup \@sanitize@url \@href}%
\providecommand \@href[1]{\@@startlink{#1}\@@href}%
\providecommand \@@href[1]{\endgroup#1\@@endlink}%
\providecommand \@sanitize@url [0]{\catcode `\\12\catcode `\$12\catcode
  `\&12\catcode `\#12\catcode `\^12\catcode `\_12\catcode `\%12\relax}%
\providecommand \@@startlink[1]{}%
\providecommand \@@endlink[0]{}%
\providecommand \url  [0]{\begingroup\@sanitize@url \@url }%
\providecommand \@url [1]{\endgroup\@href {#1}{\urlprefix }}%
\providecommand \urlprefix  [0]{URL }%
\providecommand \Eprint [0]{\href }%
\providecommand \doibase [0]{http://dx.doi.org/}%
\providecommand \selectlanguage [0]{\@gobble}%
\providecommand \bibinfo  [0]{\@secondoftwo}%
\providecommand \bibfield  [0]{\@secondoftwo}%
\providecommand \translation [1]{[#1]}%
\providecommand \BibitemOpen [0]{}%
\providecommand \bibitemStop [0]{}%
\providecommand \bibitemNoStop [0]{.\EOS\space}%
\providecommand \EOS [0]{\spacefactor3000\relax}%
\providecommand \BibitemShut  [1]{\csname bibitem#1\endcsname}%
\let\auto@bib@innerbib\@empty
\bibitem [{\citenamefont {Garc\'{\i}a~de Abajo}(2007)}]{Aba07}%
  \BibitemOpen
  \bibfield  {author} {\bibinfo {author} {\bibfnamefont {F.~J.}\ \bibnamefont
  {Garc\'{\i}a~de Abajo}},\ }\bibfield  {title} {\enquote {\bibinfo {title}
  {Colloquium: Light scattering by particle and hole arrays},}\ }\href
  {\doibase 10.1103/RevModPhys.79.1267} {\bibfield  {journal} {\bibinfo
  {journal} {Rev. Mod. Phys.}\ }\textbf {\bibinfo {volume} {79}},\ \bibinfo
  {pages} {1267--1290} (\bibinfo {year} {2007})}\BibitemShut {NoStop}%
\bibitem [{\citenamefont {Bettles}\ \emph
  {et~al.}(2016{\natexlab{a}})\citenamefont {Bettles}, \citenamefont
  {Gardiner},\ and\ \citenamefont {Adams}}]{Bet16b}%
  \BibitemOpen
  \bibfield  {author} {\bibinfo {author} {\bibfnamefont {Robert~J.}\
  \bibnamefont {Bettles}}, \bibinfo {author} {\bibfnamefont {Simon~A.}\
  \bibnamefont {Gardiner}}, \ and\ \bibinfo {author} {\bibfnamefont
  {Charles~S.}\ \bibnamefont {Adams}},\ }\bibfield  {title} {\enquote {\bibinfo
  {title} {Enhanced optical cross section via collective coupling of atomic
  dipoles in a 2d array},}\ }\href {\doibase 10.1103/PhysRevLett.116.103602}
  {\bibfield  {journal} {\bibinfo  {journal} {Phys. Rev. Lett.}\ }\textbf
  {\bibinfo {volume} {116}},\ \bibinfo {pages} {103602} (\bibinfo {year}
  {2016}{\natexlab{a}})}\BibitemShut {NoStop}%
\bibitem [{\citenamefont {Shahmoon}\ \emph {et~al.}(2017)\citenamefont
  {Shahmoon}, \citenamefont {Wild}, \citenamefont {Lukin},\ and\ \citenamefont
  {Yelin}}]{Sha17}%
  \BibitemOpen
  \bibfield  {author} {\bibinfo {author} {\bibfnamefont {Ephraim}\ \bibnamefont
  {Shahmoon}}, \bibinfo {author} {\bibfnamefont {Dominik~S.}\ \bibnamefont
  {Wild}}, \bibinfo {author} {\bibfnamefont {Mikhail~D.}\ \bibnamefont
  {Lukin}}, \ and\ \bibinfo {author} {\bibfnamefont {Susanne~F.}\ \bibnamefont
  {Yelin}},\ }\bibfield  {title} {\enquote {\bibinfo {title} {Cooperative
  resonances in light scattering from two-dimensional atomic arrays},}\ }\href
  {\doibase 10.1103/PhysRevLett.118.113601} {\bibfield  {journal} {\bibinfo
  {journal} {Phys. Rev. Lett.}\ }\textbf {\bibinfo {volume} {118}},\ \bibinfo
  {pages} {113601} (\bibinfo {year} {2017})}\BibitemShut {NoStop}%
\bibitem [{\citenamefont {Scuri}\ \emph {et~al.}(2018)\citenamefont {Scuri},
  \citenamefont {Zhou}, \citenamefont {High}, \citenamefont {Wild},
  \citenamefont {Shu}, \citenamefont {De~Greve}, \citenamefont {Jauregui},
  \citenamefont {Taniguchi}, \citenamefont {Watanabe}, \citenamefont {Kim},
  \citenamefont {Lukin},\ and\ \citenamefont {Park}}]{Scu18}%
  \BibitemOpen
  \bibfield  {author} {\bibinfo {author} {\bibfnamefont {Giovanni}\
  \bibnamefont {Scuri}}, \bibinfo {author} {\bibfnamefont {You}\ \bibnamefont
  {Zhou}}, \bibinfo {author} {\bibfnamefont {Alexander~A.}\ \bibnamefont
  {High}}, \bibinfo {author} {\bibfnamefont {Dominik~S.}\ \bibnamefont {Wild}},
  \bibinfo {author} {\bibfnamefont {Chi}\ \bibnamefont {Shu}}, \bibinfo
  {author} {\bibfnamefont {Kristiaan}\ \bibnamefont {De~Greve}}, \bibinfo
  {author} {\bibfnamefont {Luis~A.}\ \bibnamefont {Jauregui}}, \bibinfo
  {author} {\bibfnamefont {Takashi}\ \bibnamefont {Taniguchi}}, \bibinfo
  {author} {\bibfnamefont {Kenji}\ \bibnamefont {Watanabe}}, \bibinfo {author}
  {\bibfnamefont {Philip}\ \bibnamefont {Kim}}, \bibinfo {author}
  {\bibfnamefont {Mikhail~D.}\ \bibnamefont {Lukin}}, \ and\ \bibinfo {author}
  {\bibfnamefont {Hongkun}\ \bibnamefont {Park}},\ }\bibfield  {title}
  {\enquote {\bibinfo {title} {{Large Excitonic Reflectivity of Monolayer
  ${\mathrm{MoSe}}_{2}$ Encapsulated in Hexagonal Boron Nitride}},}\ }\href
  {\doibase 10.1103/PhysRevLett.120.037402} {\bibfield  {journal} {\bibinfo
  {journal} {Phys. Rev. Lett.}\ }\textbf {\bibinfo {volume} {120}},\ \bibinfo
  {pages} {037402} (\bibinfo {year} {2018})}\BibitemShut {NoStop}%
\bibitem [{\citenamefont {Back}\ \emph {et~al.}(2018)\citenamefont {Back},
  \citenamefont {Zeytinoglu}, \citenamefont {Ijaz}, \citenamefont {Kroner},\
  and\ \citenamefont {Imamoglu}}]{Bac18}%
  \BibitemOpen
  \bibfield  {author} {\bibinfo {author} {\bibfnamefont {Patrick}\ \bibnamefont
  {Back}}, \bibinfo {author} {\bibfnamefont {Sina}\ \bibnamefont {Zeytinoglu}},
  \bibinfo {author} {\bibfnamefont {Aroosa}\ \bibnamefont {Ijaz}}, \bibinfo
  {author} {\bibfnamefont {Martin}\ \bibnamefont {Kroner}}, \ and\ \bibinfo
  {author} {\bibfnamefont {Atac}\ \bibnamefont {Imamoglu}},\ }\bibfield
  {title} {\enquote {\bibinfo {title} {Realization of an electrically tunable
  narrow-bandwidth atomically thin mirror using monolayer
  {${\mathrm{MoSe}}_{2}$}},}\ }\href {\doibase 10.1103/PhysRevLett.120.037401}
  {\bibfield  {journal} {\bibinfo  {journal} {Phys. Rev. Lett.}\ }\textbf
  {\bibinfo {volume} {120}},\ \bibinfo {pages} {037401} (\bibinfo {year}
  {2018})}\BibitemShut {NoStop}%
\bibitem [{\citenamefont {Bender}\ \emph {et~al.}(2010)\citenamefont {Bender},
  \citenamefont {Stehle}, \citenamefont {Slama}, \citenamefont {Kaiser},
  \citenamefont {Piovella}, \citenamefont {Zimmermann},\ and\ \citenamefont
  {Courteille}}]{Ben10}%
  \BibitemOpen
  \bibfield  {author} {\bibinfo {author} {\bibfnamefont {H.}~\bibnamefont
  {Bender}}, \bibinfo {author} {\bibfnamefont {C.}~\bibnamefont {Stehle}},
  \bibinfo {author} {\bibfnamefont {S.}~\bibnamefont {Slama}}, \bibinfo
  {author} {\bibfnamefont {R.}~\bibnamefont {Kaiser}}, \bibinfo {author}
  {\bibfnamefont {N.}~\bibnamefont {Piovella}}, \bibinfo {author}
  {\bibfnamefont {C.}~\bibnamefont {Zimmermann}}, \ and\ \bibinfo {author}
  {\bibfnamefont {Ph.~W.}\ \bibnamefont {Courteille}},\ }\bibfield  {title}
  {\enquote {\bibinfo {title} {Observation of cooperative mie scattering from
  an ultracold atomic cloud},}\ }\href {\doibase 10.1103/PhysRevA.82.011404}
  {\bibfield  {journal} {\bibinfo  {journal} {Phys. Rev. A}\ }\textbf {\bibinfo
  {volume} {82}},\ \bibinfo {pages} {011404} (\bibinfo {year}
  {2010})}\BibitemShut {NoStop}%
\bibitem [{\citenamefont {Pellegrino}\ \emph {et~al.}(2014)\citenamefont
  {Pellegrino}, \citenamefont {Bourgain}, \citenamefont {Jennewein},
  \citenamefont {Sortais}, \citenamefont {Browaeys}, \citenamefont {Jenkins},\
  and\ \citenamefont {Ruostekoski}}]{Pel14}%
  \BibitemOpen
  \bibfield  {author} {\bibinfo {author} {\bibfnamefont {J.}~\bibnamefont
  {Pellegrino}}, \bibinfo {author} {\bibfnamefont {R.}~\bibnamefont
  {Bourgain}}, \bibinfo {author} {\bibfnamefont {S.}~\bibnamefont {Jennewein}},
  \bibinfo {author} {\bibfnamefont {Y.~R.~P.}\ \bibnamefont {Sortais}},
  \bibinfo {author} {\bibfnamefont {A.}~\bibnamefont {Browaeys}}, \bibinfo
  {author} {\bibfnamefont {S.~D.}\ \bibnamefont {Jenkins}}, \ and\ \bibinfo
  {author} {\bibfnamefont {J.}~\bibnamefont {Ruostekoski}},\ }\bibfield
  {title} {\enquote {\bibinfo {title} {Observation of suppression of light
  scattering induced by dipole-dipole interactions in a cold-atom ensemble},}\
  }\href {\doibase 10.1103/PhysRevLett.113.133602} {\bibfield  {journal}
  {\bibinfo  {journal} {Phys. Rev. Lett.}\ }\textbf {\bibinfo {volume} {113}},\
  \bibinfo {pages} {133602} (\bibinfo {year} {2014})}\BibitemShut {NoStop}%
\bibitem [{\citenamefont {Guerin}\ \emph {et~al.}(2016)\citenamefont {Guerin},
  \citenamefont {Ara\'ujo},\ and\ \citenamefont {Kaiser}}]{Gue16}%
  \BibitemOpen
  \bibfield  {author} {\bibinfo {author} {\bibfnamefont {William}\ \bibnamefont
  {Guerin}}, \bibinfo {author} {\bibfnamefont {Michelle~O.}\ \bibnamefont
  {Ara\'ujo}}, \ and\ \bibinfo {author} {\bibfnamefont {Robin}\ \bibnamefont
  {Kaiser}},\ }\bibfield  {title} {\enquote {\bibinfo {title} {Subradiance in a
  large cloud of cold atoms},}\ }\href {\doibase
  10.1103/PhysRevLett.116.083601} {\bibfield  {journal} {\bibinfo  {journal}
  {Phys. Rev. Lett.}\ }\textbf {\bibinfo {volume} {116}},\ \bibinfo {pages}
  {083601} (\bibinfo {year} {2016})}\BibitemShut {NoStop}%
\bibitem [{\citenamefont {Ara\'ujo}\ \emph {et~al.}(2016)\citenamefont
  {Ara\'ujo}, \citenamefont {Kre\ifmmode \check{s}\else
  \v{s}\fi{}i\ifmmode~\acute{c}\else \'{c}\fi{}}, \citenamefont {Kaiser},\ and\
  \citenamefont {Guerin}}]{Ara16}%
  \BibitemOpen
  \bibfield  {author} {\bibinfo {author} {\bibfnamefont {Michelle~O.}\
  \bibnamefont {Ara\'ujo}}, \bibinfo {author} {\bibfnamefont {Ivor}\
  \bibnamefont {Kre\ifmmode \check{s}\else \v{s}\fi{}i\ifmmode~\acute{c}\else
  \'{c}\fi{}}}, \bibinfo {author} {\bibfnamefont {Robin}\ \bibnamefont
  {Kaiser}}, \ and\ \bibinfo {author} {\bibfnamefont {William}\ \bibnamefont
  {Guerin}},\ }\bibfield  {title} {\enquote {\bibinfo {title} {Superradiance in
  a large and dilute cloud of cold atoms in the linear-optics regime},}\ }\href
  {\doibase 10.1103/PhysRevLett.117.073002} {\bibfield  {journal} {\bibinfo
  {journal} {Phys. Rev. Lett.}\ }\textbf {\bibinfo {volume} {117}},\ \bibinfo
  {pages} {073002} (\bibinfo {year} {2016})}\BibitemShut {NoStop}%
\bibitem [{\citenamefont {Roof}\ \emph {et~al.}(2016)\citenamefont {Roof},
  \citenamefont {Kemp}, \citenamefont {Havey},\ and\ \citenamefont
  {Sokolov}}]{Roo16}%
  \BibitemOpen
  \bibfield  {author} {\bibinfo {author} {\bibfnamefont {S.~J.}\ \bibnamefont
  {Roof}}, \bibinfo {author} {\bibfnamefont {K.~J.}\ \bibnamefont {Kemp}},
  \bibinfo {author} {\bibfnamefont {M.~D.}\ \bibnamefont {Havey}}, \ and\
  \bibinfo {author} {\bibfnamefont {I.~M.}\ \bibnamefont {Sokolov}},\
  }\bibfield  {title} {\enquote {\bibinfo {title} {Observation of single-photon
  superradiance and the cooperative lamb shift in an extended sample of cold
  atoms},}\ }\href {\doibase 10.1103/PhysRevLett.117.073003} {\bibfield
  {journal} {\bibinfo  {journal} {Phys. Rev. Lett.}\ }\textbf {\bibinfo
  {volume} {117}},\ \bibinfo {pages} {073003} (\bibinfo {year}
  {2016})}\BibitemShut {NoStop}%
\bibitem [{\citenamefont {Jennewein}\ \emph {et~al.}(2016)\citenamefont
  {Jennewein}, \citenamefont {Besbes}, \citenamefont {Schilder}, \citenamefont
  {Jenkins}, \citenamefont {Sauvan}, \citenamefont {Ruostekoski}, \citenamefont
  {Greffet}, \citenamefont {Sortais},\ and\ \citenamefont {Browaeys}}]{Jen16}%
  \BibitemOpen
  \bibfield  {author} {\bibinfo {author} {\bibfnamefont {S.}~\bibnamefont
  {Jennewein}}, \bibinfo {author} {\bibfnamefont {M.}~\bibnamefont {Besbes}},
  \bibinfo {author} {\bibfnamefont {N.~J.}\ \bibnamefont {Schilder}}, \bibinfo
  {author} {\bibfnamefont {S.~D.}\ \bibnamefont {Jenkins}}, \bibinfo {author}
  {\bibfnamefont {C.}~\bibnamefont {Sauvan}}, \bibinfo {author} {\bibfnamefont
  {J.}~\bibnamefont {Ruostekoski}}, \bibinfo {author} {\bibfnamefont {J.-J.}\
  \bibnamefont {Greffet}}, \bibinfo {author} {\bibfnamefont {Y.~R.~P.}\
  \bibnamefont {Sortais}}, \ and\ \bibinfo {author} {\bibfnamefont
  {A.}~\bibnamefont {Browaeys}},\ }\bibfield  {title} {\enquote {\bibinfo
  {title} {{Coherent Scattering of Near-Resonant Light by a Dense Microscopic
  Cold Atomic Cloud}},}\ }\href {\doibase 10.1103/PhysRevLett.116.233601}
  {\bibfield  {journal} {\bibinfo  {journal} {Phys. Rev. Lett.}\ }\textbf
  {\bibinfo {volume} {116}},\ \bibinfo {pages} {233601} (\bibinfo {year}
  {2016})}\BibitemShut {NoStop}%
\bibitem [{\citenamefont {Guerin}\ \emph {et~al.}(2017)\citenamefont {Guerin},
  \citenamefont {Rouabah},\ and\ \citenamefont {Kaiser}}]{Gue17}%
  \BibitemOpen
  \bibfield  {author} {\bibinfo {author} {\bibfnamefont {W.}~\bibnamefont
  {Guerin}}, \bibinfo {author} {\bibfnamefont {M.T.}\ \bibnamefont {Rouabah}},
  \ and\ \bibinfo {author} {\bibfnamefont {R.}~\bibnamefont {Kaiser}},\
  }\bibfield  {title} {\enquote {\bibinfo {title} {Light interacting with
  atomic ensembles: collective, cooperative and mesoscopic effects},}\ }\href
  {\doibase 10.1080/09500340.2016.1215564} {\bibfield  {journal} {\bibinfo
  {journal} {Journal of Modern Optics}\ }\textbf {\bibinfo {volume} {64}},\
  \bibinfo {pages} {895--907} (\bibinfo {year} {2017})}\BibitemShut {NoStop}%
\bibitem [{\citenamefont {Bromley}\ \emph {et~al.}(2016)\citenamefont
  {Bromley}, \citenamefont {Zhu}, \citenamefont {Bishof}, \citenamefont
  {Zhang}, \citenamefont {Bothwell}, \citenamefont {Schachenmayer},
  \citenamefont {Nicholson}, \citenamefont {Kaiser}, \citenamefont {Yelin},
  \citenamefont {Lukin}, \citenamefont {Rey},\ and\ \citenamefont
  {Ye}}]{Brom16}%
  \BibitemOpen
  \bibfield  {author} {\bibinfo {author} {\bibfnamefont {S.~L.}\ \bibnamefont
  {Bromley}}, \bibinfo {author} {\bibfnamefont {B.}~\bibnamefont {Zhu}},
  \bibinfo {author} {\bibfnamefont {M.}~\bibnamefont {Bishof}}, \bibinfo
  {author} {\bibfnamefont {X.}~\bibnamefont {Zhang}}, \bibinfo {author}
  {\bibfnamefont {T.}~\bibnamefont {Bothwell}}, \bibinfo {author}
  {\bibfnamefont {J.}~\bibnamefont {Schachenmayer}}, \bibinfo {author}
  {\bibfnamefont {T.~L.}\ \bibnamefont {Nicholson}}, \bibinfo {author}
  {\bibfnamefont {R.}~\bibnamefont {Kaiser}}, \bibinfo {author} {\bibfnamefont
  {S.~F.}\ \bibnamefont {Yelin}}, \bibinfo {author} {\bibfnamefont {M.~D.}\
  \bibnamefont {Lukin}}, \bibinfo {author} {\bibfnamefont {A.~M.}\ \bibnamefont
  {Rey}}, \ and\ \bibinfo {author} {\bibfnamefont {J.}~\bibnamefont {Ye}},\
  }\bibfield  {title} {\enquote {\bibinfo {title} {Collective atomic scattering
  and motional effects in a dense coherent medium},}\ }\href {\doibase
  10.1038/ncomms11039} {\bibfield  {journal} {\bibinfo  {journal} {Nature
  Communications}\ }\textbf {\bibinfo {volume} {7}},\ \bibinfo {pages} {11039}
  (\bibinfo {year} {2016})}\BibitemShut {NoStop}%
\bibitem [{\citenamefont {Corman}\ \emph {et~al.}(2017)\citenamefont {Corman},
  \citenamefont {Ville}, \citenamefont {Saint-Jalm}, \citenamefont
  {Aidelsburger}, \citenamefont {Bienaim\'e}, \citenamefont {Nascimb\`ene},
  \citenamefont {Dalibard},\ and\ \citenamefont {Beugnon}}]{Cor17}%
  \BibitemOpen
  \bibfield  {author} {\bibinfo {author} {\bibfnamefont {L.}~\bibnamefont
  {Corman}}, \bibinfo {author} {\bibfnamefont {J.~L.}\ \bibnamefont {Ville}},
  \bibinfo {author} {\bibfnamefont {R.}~\bibnamefont {Saint-Jalm}}, \bibinfo
  {author} {\bibfnamefont {M.}~\bibnamefont {Aidelsburger}}, \bibinfo {author}
  {\bibfnamefont {T.}~\bibnamefont {Bienaim\'e}}, \bibinfo {author}
  {\bibfnamefont {S.}~\bibnamefont {Nascimb\`ene}}, \bibinfo {author}
  {\bibfnamefont {J.}~\bibnamefont {Dalibard}}, \ and\ \bibinfo {author}
  {\bibfnamefont {J.}~\bibnamefont {Beugnon}},\ }\bibfield  {title} {\enquote
  {\bibinfo {title} {Transmission of near-resonant light through a dense slab
  of cold atoms},}\ }\href {\doibase 10.1103/PhysRevA.96.053629} {\bibfield
  {journal} {\bibinfo  {journal} {Phys. Rev. A}\ }\textbf {\bibinfo {volume}
  {96}},\ \bibinfo {pages} {053629} (\bibinfo {year} {2017})}\BibitemShut
  {NoStop}%
\bibitem [{\citenamefont {Jennewein}\ \emph {et~al.}(2018)\citenamefont
  {Jennewein}, \citenamefont {Brossard}, \citenamefont {Sortais}, \citenamefont
  {Browaeys}, \citenamefont {Cheinet}, \citenamefont {Robert},\ and\
  \citenamefont {Pillet}}]{Jen18}%
  \BibitemOpen
  \bibfield  {author} {\bibinfo {author} {\bibfnamefont {Stephan}\ \bibnamefont
  {Jennewein}}, \bibinfo {author} {\bibfnamefont {Ludovic}\ \bibnamefont
  {Brossard}}, \bibinfo {author} {\bibfnamefont {Yvan R.~P.}\ \bibnamefont
  {Sortais}}, \bibinfo {author} {\bibfnamefont {Antoine}\ \bibnamefont
  {Browaeys}}, \bibinfo {author} {\bibfnamefont {Patrick}\ \bibnamefont
  {Cheinet}}, \bibinfo {author} {\bibfnamefont {Jacques}\ \bibnamefont
  {Robert}}, \ and\ \bibinfo {author} {\bibfnamefont {Pierre}\ \bibnamefont
  {Pillet}},\ }\bibfield  {title} {\enquote {\bibinfo {title} {{Coherent
  scattering of near-resonant light by a dense, microscopic cloud of cold
  two-level atoms: Experiment versus theory}},}\ }\href {\doibase
  10.1103/PhysRevA.97.053816} {\bibfield  {journal} {\bibinfo  {journal} {Phys.
  Rev. A}\ }\textbf {\bibinfo {volume} {97}},\ \bibinfo {pages} {053816}
  (\bibinfo {year} {2018})}\BibitemShut {NoStop}%
\bibitem [{\citenamefont {Facchinetti}\ and\ \citenamefont
  {Ruostekoski}(2018)}]{Fac18}%
  \BibitemOpen
  \bibfield  {author} {\bibinfo {author} {\bibfnamefont {G.}~\bibnamefont
  {Facchinetti}}\ and\ \bibinfo {author} {\bibfnamefont {J.}~\bibnamefont
  {Ruostekoski}},\ }\bibfield  {title} {\enquote {\bibinfo {title} {Interaction
  of light with planar lattices of atoms: Reflection, transmission, and
  cooperative magnetometry},}\ }\href {\doibase 10.1103/PhysRevA.97.023833}
  {\bibfield  {journal} {\bibinfo  {journal} {Phys. Rev. A}\ }\textbf {\bibinfo
  {volume} {97}},\ \bibinfo {pages} {023833} (\bibinfo {year}
  {2018})}\BibitemShut {NoStop}%
\bibitem [{\citenamefont {Rui}\ \emph {et~al.}(2020)\citenamefont {Rui},
  \citenamefont {Wei}, \citenamefont {Rubio-Abadal}, \citenamefont {Hollerith},
  \citenamefont {Zeiher}, \citenamefont {Stamper-Kurn}, \citenamefont {Gross},\
  and\ \citenamefont {Bloch}}]{Rui20}%
  \BibitemOpen
  \bibfield  {author} {\bibinfo {author} {\bibfnamefont {Jun}\ \bibnamefont
  {Rui}}, \bibinfo {author} {\bibfnamefont {David}\ \bibnamefont {Wei}},
  \bibinfo {author} {\bibfnamefont {Antonio}\ \bibnamefont {Rubio-Abadal}},
  \bibinfo {author} {\bibfnamefont {Simon}\ \bibnamefont {Hollerith}}, \bibinfo
  {author} {\bibfnamefont {Johannes}\ \bibnamefont {Zeiher}}, \bibinfo {author}
  {\bibfnamefont {Dan~M.}\ \bibnamefont {Stamper-Kurn}}, \bibinfo {author}
  {\bibfnamefont {Christian}\ \bibnamefont {Gross}}, \ and\ \bibinfo {author}
  {\bibfnamefont {Immanuel}\ \bibnamefont {Bloch}},\ }\href@noop {} {\enquote
  {\bibinfo {title} {A subradiant optical mirror formed by a single structured
  atomic layer},}\ } (\bibinfo {year} {2020}),\ \Eprint
  {http://arxiv.org/abs/2001.00795} {arXiv:2001.00795 [quant-ph]} \BibitemShut
  {NoStop}%
\bibitem [{\citenamefont {Plankensteiner}\ \emph {et~al.}(2015)\citenamefont
  {Plankensteiner}, \citenamefont {Ostermann}, \citenamefont {Ritsch},\ and\
  \citenamefont {Genes}}]{Plan15}%
  \BibitemOpen
  \bibfield  {author} {\bibinfo {author} {\bibfnamefont {D.}~\bibnamefont
  {Plankensteiner}}, \bibinfo {author} {\bibfnamefont {L.}~\bibnamefont
  {Ostermann}}, \bibinfo {author} {\bibfnamefont {H.}~\bibnamefont {Ritsch}}, \
  and\ \bibinfo {author} {\bibfnamefont {C.}~\bibnamefont {Genes}},\ }\bibfield
   {title} {\enquote {\bibinfo {title} {Selective protected state preparation
  of coupled dissipative quantum emitters},}\ }\href {\doibase
  10.1038/srep16231} {\bibfield  {journal} {\bibinfo  {journal} {Scientific
  Reports}\ }\textbf {\bibinfo {volume} {5}},\ \bibinfo {pages} {16231}
  (\bibinfo {year} {2015})}\BibitemShut {NoStop}%
\bibitem [{\citenamefont {Chui}\ \emph {et~al.}(2015)\citenamefont {Chui},
  \citenamefont {Du},\ and\ \citenamefont {Jo}}]{Chu15}%
  \BibitemOpen
  \bibfield  {author} {\bibinfo {author} {\bibfnamefont {Siu-Tat}\ \bibnamefont
  {Chui}}, \bibinfo {author} {\bibfnamefont {Shengwang}\ \bibnamefont {Du}}, \
  and\ \bibinfo {author} {\bibfnamefont {Gyu-Boong}\ \bibnamefont {Jo}},\
  }\bibfield  {title} {\enquote {\bibinfo {title} {Subwavelength transportation
  of light with atomic resonances},}\ }\href {\doibase
  10.1103/PhysRevA.92.053826} {\bibfield  {journal} {\bibinfo  {journal} {Phys.
  Rev. A}\ }\textbf {\bibinfo {volume} {92}},\ \bibinfo {pages} {053826}
  (\bibinfo {year} {2015})}\BibitemShut {NoStop}%
\bibitem [{\citenamefont {Needham}\ \emph {et~al.}(2019)\citenamefont
  {Needham}, \citenamefont {Lesanovsky},\ and\ \citenamefont {Olmos}}]{Nee19}%
  \BibitemOpen
  \bibfield  {author} {\bibinfo {author} {\bibfnamefont {Jemma~A}\ \bibnamefont
  {Needham}}, \bibinfo {author} {\bibfnamefont {Igor}\ \bibnamefont
  {Lesanovsky}}, \ and\ \bibinfo {author} {\bibfnamefont {Beatriz}\
  \bibnamefont {Olmos}},\ }\bibfield  {title} {\enquote {\bibinfo {title}
  {Subradiance-protected excitation transport},}\ }\href {\doibase
  10.1088/1367-2630/ab31e8} {\bibfield  {journal} {\bibinfo  {journal} {New
  Journal of Physics}\ }\textbf {\bibinfo {volume} {21}},\ \bibinfo {pages}
  {073061} (\bibinfo {year} {2019})}\BibitemShut {NoStop}%
\bibitem [{\citenamefont {Bettles}\ \emph
  {et~al.}(2016{\natexlab{b}})\citenamefont {Bettles}, \citenamefont
  {Gardiner},\ and\ \citenamefont {Adams}}]{Bet16a}%
  \BibitemOpen
  \bibfield  {author} {\bibinfo {author} {\bibfnamefont {Robert~J.}\
  \bibnamefont {Bettles}}, \bibinfo {author} {\bibfnamefont {Simon~A.}\
  \bibnamefont {Gardiner}}, \ and\ \bibinfo {author} {\bibfnamefont
  {Charles~S.}\ \bibnamefont {Adams}},\ }\bibfield  {title} {\enquote {\bibinfo
  {title} {Cooperative eigenmodes and scattering in one-dimensional atomic
  arrays},}\ }\href {\doibase 10.1103/PhysRevA.94.043844} {\bibfield  {journal}
  {\bibinfo  {journal} {Phys. Rev. A}\ }\textbf {\bibinfo {volume} {94}},\
  \bibinfo {pages} {043844} (\bibinfo {year} {2016}{\natexlab{b}})}\BibitemShut
  {NoStop}%
\bibitem [{\citenamefont {Asenjo-Garcia}\ \emph {et~al.}(2017)\citenamefont
  {Asenjo-Garcia}, \citenamefont {Moreno-Cardoner}, \citenamefont {Albrecht},
  \citenamefont {Kimble},\ and\ \citenamefont {Chang}}]{Ase17}%
  \BibitemOpen
  \bibfield  {author} {\bibinfo {author} {\bibfnamefont {A.}~\bibnamefont
  {Asenjo-Garcia}}, \bibinfo {author} {\bibfnamefont {M.}~\bibnamefont
  {Moreno-Cardoner}}, \bibinfo {author} {\bibfnamefont {A.}~\bibnamefont
  {Albrecht}}, \bibinfo {author} {\bibfnamefont {H.~J.}\ \bibnamefont
  {Kimble}}, \ and\ \bibinfo {author} {\bibfnamefont {D.~E.}\ \bibnamefont
  {Chang}},\ }\bibfield  {title} {\enquote {\bibinfo {title} {Exponential
  improvement in photon storage fidelities using subradiance and ``selective
  radiance'' in atomic arrays},}\ }\href {\doibase 10.1103/PhysRevX.7.031024}
  {\bibfield  {journal} {\bibinfo  {journal} {Phys. Rev. X}\ }\textbf {\bibinfo
  {volume} {7}},\ \bibinfo {pages} {031024} (\bibinfo {year}
  {2017})}\BibitemShut {NoStop}%
\bibitem [{\citenamefont {Javanainen}\ \emph {et~al.}(2014)\citenamefont
  {Javanainen}, \citenamefont {Ruostekoski}, \citenamefont {Li},\ and\
  \citenamefont {Yoo}}]{Jav14}%
  \BibitemOpen
  \bibfield  {author} {\bibinfo {author} {\bibfnamefont {Juha}\ \bibnamefont
  {Javanainen}}, \bibinfo {author} {\bibfnamefont {Janne}\ \bibnamefont
  {Ruostekoski}}, \bibinfo {author} {\bibfnamefont {Yi}~\bibnamefont {Li}}, \
  and\ \bibinfo {author} {\bibfnamefont {Sung-Mi}\ \bibnamefont {Yoo}},\
  }\bibfield  {title} {\enquote {\bibinfo {title} {Shifts of a resonance line
  in a dense atomic sample},}\ }\href {\doibase 10.1103/PhysRevLett.112.113603}
  {\bibfield  {journal} {\bibinfo  {journal} {Phys. Rev. Lett.}\ }\textbf
  {\bibinfo {volume} {112}},\ \bibinfo {pages} {113603} (\bibinfo {year}
  {2014})}\BibitemShut {NoStop}%
\bibitem [{\citenamefont {Jones}\ \emph {et~al.}(2017)\citenamefont {Jones},
  \citenamefont {Saint},\ and\ \citenamefont {Olmos}}]{Jon16}%
  \BibitemOpen
  \bibfield  {author} {\bibinfo {author} {\bibfnamefont {Ryan}\ \bibnamefont
  {Jones}}, \bibinfo {author} {\bibfnamefont {Reece}\ \bibnamefont {Saint}}, \
  and\ \bibinfo {author} {\bibfnamefont {Beatriz}\ \bibnamefont {Olmos}},\
  }\bibfield  {title} {\enquote {\bibinfo {title} {Far-field resonance
  fluorescence from a dipole-interacting laser-driven cold atomic gas},}\
  }\href {\doibase 10.1088/1361-6455/50/1/014004} {\bibfield  {journal}
  {\bibinfo  {journal} {Journal of Physics B: Atomic, Molecular and Optical
  Physics}\ }\textbf {\bibinfo {volume} {50}},\ \bibinfo {pages} {014004}
  (\bibinfo {year} {2017})}\BibitemShut {NoStop}%
\bibitem [{\citenamefont {Williamson}\ and\ \citenamefont
  {Ruostekoski}(2020)}]{Wil20}%
  \BibitemOpen
  \bibfield  {author} {\bibinfo {author} {\bibfnamefont {L.~A.}\ \bibnamefont
  {Williamson}}\ and\ \bibinfo {author} {\bibfnamefont {J.}~\bibnamefont
  {Ruostekoski}},\ }\href@noop {} {\enquote {\bibinfo {title} {Optical response
  of atom chains beyond the limit of low light intensity: The validity of the
  linear classical oscillator model},}\ } (\bibinfo {year} {2020}),\ \Eprint
  {http://arxiv.org/abs/2002.01417} {arXiv:2002.01417 [physics.atom-ph]}
  \BibitemShut {NoStop}%
\bibitem [{\citenamefont {Yu}\ \emph {et~al.}(2014)\citenamefont {Yu},
  \citenamefont {Hood}, \citenamefont {Muniz}, \citenamefont {Martin},
  \citenamefont {Norte}, \citenamefont {Hung}, \citenamefont {Meenehan},
  \citenamefont {Cohen}, \citenamefont {Painter},\ and\ \citenamefont
  {Kimble}}]{Yu14}%
  \BibitemOpen
  \bibfield  {author} {\bibinfo {author} {\bibfnamefont {S.-P.}\ \bibnamefont
  {Yu}}, \bibinfo {author} {\bibfnamefont {J.~D.}\ \bibnamefont {Hood}},
  \bibinfo {author} {\bibfnamefont {J.~A.}\ \bibnamefont {Muniz}}, \bibinfo
  {author} {\bibfnamefont {M.~J.}\ \bibnamefont {Martin}}, \bibinfo {author}
  {\bibfnamefont {Richard}\ \bibnamefont {Norte}}, \bibinfo {author}
  {\bibfnamefont {C.-L.}\ \bibnamefont {Hung}}, \bibinfo {author}
  {\bibfnamefont {SeÃ¡n~M.}\ \bibnamefont {Meenehan}}, \bibinfo {author}
  {\bibfnamefont {Justin~D.}\ \bibnamefont {Cohen}}, \bibinfo {author}
  {\bibfnamefont {Oskar}\ \bibnamefont {Painter}}, \ and\ \bibinfo {author}
  {\bibfnamefont {H.~J.}\ \bibnamefont {Kimble}},\ }\bibfield  {title}
  {\enquote {\bibinfo {title} {Nanowire photonic crystal waveguides for
  single-atom trapping and strong light-matter interactions},}\ }\href
  {\doibase 10.1063/1.4868975} {\bibfield  {journal} {\bibinfo  {journal}
  {Applied Physics Letters}\ }\textbf {\bibinfo {volume} {104}},\ \bibinfo
  {pages} {111103} (\bibinfo {year} {2014})}\BibitemShut {NoStop}%
\bibitem [{\citenamefont {Vetsch}\ \emph {et~al.}(2010)\citenamefont {Vetsch},
  \citenamefont {Reitz}, \citenamefont {Sagu\'e}, \citenamefont {Schmidt},
  \citenamefont {Dawkins},\ and\ \citenamefont {Rauschenbeutel}}]{Vet10}%
  \BibitemOpen
  \bibfield  {author} {\bibinfo {author} {\bibfnamefont {E.}~\bibnamefont
  {Vetsch}}, \bibinfo {author} {\bibfnamefont {D.}~\bibnamefont {Reitz}},
  \bibinfo {author} {\bibfnamefont {G.}~\bibnamefont {Sagu\'e}}, \bibinfo
  {author} {\bibfnamefont {R.}~\bibnamefont {Schmidt}}, \bibinfo {author}
  {\bibfnamefont {S.~T.}\ \bibnamefont {Dawkins}}, \ and\ \bibinfo {author}
  {\bibfnamefont {A.}~\bibnamefont {Rauschenbeutel}},\ }\bibfield  {title}
  {\enquote {\bibinfo {title} {Optical interface created by laser-cooled atoms
  trapped in the evanescent field surrounding an optical nanofiber},}\ }\href
  {\doibase 10.1103/PhysRevLett.104.203603} {\bibfield  {journal} {\bibinfo
  {journal} {Phys. Rev. Lett.}\ }\textbf {\bibinfo {volume} {104}},\ \bibinfo
  {pages} {203603} (\bibinfo {year} {2010})}\BibitemShut {NoStop}%
\bibitem [{\citenamefont {Solano}\ \emph {et~al.}(2017)\citenamefont {Solano},
  \citenamefont {Barberis-Blostein}, \citenamefont {Fatemi}, \citenamefont
  {Orozco},\ and\ \citenamefont {Rolston}}]{Sol17}%
  \BibitemOpen
  \bibfield  {author} {\bibinfo {author} {\bibfnamefont {P.}~\bibnamefont
  {Solano}}, \bibinfo {author} {\bibfnamefont {P.}~\bibnamefont
  {Barberis-Blostein}}, \bibinfo {author} {\bibfnamefont {F.~K.}\ \bibnamefont
  {Fatemi}}, \bibinfo {author} {\bibfnamefont {L.~A.}\ \bibnamefont {Orozco}},
  \ and\ \bibinfo {author} {\bibfnamefont {S.~L.}\ \bibnamefont {Rolston}},\
  }\bibfield  {title} {\enquote {\bibinfo {title} {Super-radiance reveals
  infinite-range dipole interactions through a nanofiber},}\ }\href {\doibase
  10.1038/s41467-017-01994-3} {\bibfield  {journal} {\bibinfo  {journal}
  {Nature Communications}\ }\textbf {\bibinfo {volume} {8}},\ \bibinfo {pages}
  {1857} (\bibinfo {year} {2017})}\BibitemShut {NoStop}%
\bibitem [{\citenamefont {Corzo}\ \emph {et~al.}(2019)\citenamefont {Corzo},
  \citenamefont {Raskop}, \citenamefont {Chandra}, \citenamefont {Sheremet},
  \citenamefont {Gouraud},\ and\ \citenamefont {Laurat}}]{Cor19}%
  \BibitemOpen
  \bibfield  {author} {\bibinfo {author} {\bibfnamefont {Neil~V.}\ \bibnamefont
  {Corzo}}, \bibinfo {author} {\bibfnamefont {J{\'e}r{\'e}my}\ \bibnamefont
  {Raskop}}, \bibinfo {author} {\bibfnamefont {Aveek}\ \bibnamefont {Chandra}},
  \bibinfo {author} {\bibfnamefont {Alexandra~S.}\ \bibnamefont {Sheremet}},
  \bibinfo {author} {\bibfnamefont {Baptiste}\ \bibnamefont {Gouraud}}, \ and\
  \bibinfo {author} {\bibfnamefont {Julien}\ \bibnamefont {Laurat}},\
  }\bibfield  {title} {\enquote {\bibinfo {title} {Waveguide-coupled single
  collective excitation of atomic arrays},}\ }\href {\doibase
  10.1038/s41586-019-0902-3} {\bibfield  {journal} {\bibinfo  {journal}
  {Nature}\ }\textbf {\bibinfo {volume} {566}},\ \bibinfo {pages} {359--362}
  (\bibinfo {year} {2019})}\BibitemShut {NoStop}%
\bibitem [{\citenamefont {Prasad}\ \emph {et~al.}(2019)\citenamefont {Prasad},
  \citenamefont {Hinney}, \citenamefont {Mahmoodian}, \citenamefont {Hammerer},
  \citenamefont {Rind}, \citenamefont {Schneeweiss}, \citenamefont {Sørensen},
  \citenamefont {Volz},\ and\ \citenamefont {Rauschenbeutel}}]{Pra19}%
  \BibitemOpen
  \bibfield  {author} {\bibinfo {author} {\bibfnamefont {Adarsh~S.}\
  \bibnamefont {Prasad}}, \bibinfo {author} {\bibfnamefont {Jakob}\
  \bibnamefont {Hinney}}, \bibinfo {author} {\bibfnamefont {Sahand}\
  \bibnamefont {Mahmoodian}}, \bibinfo {author} {\bibfnamefont {Klemens}\
  \bibnamefont {Hammerer}}, \bibinfo {author} {\bibfnamefont {Samuel}\
  \bibnamefont {Rind}}, \bibinfo {author} {\bibfnamefont {Philipp}\
  \bibnamefont {Schneeweiss}}, \bibinfo {author} {\bibfnamefont {Anders~S.}\
  \bibnamefont {Sørensen}}, \bibinfo {author} {\bibfnamefont {Jürgen}\
  \bibnamefont {Volz}}, \ and\ \bibinfo {author} {\bibfnamefont {Arno}\
  \bibnamefont {Rauschenbeutel}},\ }\href@noop {} {\enquote {\bibinfo {title}
  {Correlating photons using the collective nonlinear response of atoms weakly
  coupled to an optical mode},}\ } (\bibinfo {year} {2019}),\ \Eprint
  {http://arxiv.org/abs/1911.09701} {arXiv:1911.09701 [quant-ph]} \BibitemShut
  {NoStop}%
\bibitem [{\citenamefont {Meir}\ \emph {et~al.}(2014)\citenamefont {Meir},
  \citenamefont {Schwartz}, \citenamefont {Shahmoon}, \citenamefont {Oron},\
  and\ \citenamefont {Ozeri}}]{Mei14}%
  \BibitemOpen
  \bibfield  {author} {\bibinfo {author} {\bibfnamefont {Z.}~\bibnamefont
  {Meir}}, \bibinfo {author} {\bibfnamefont {O.}~\bibnamefont {Schwartz}},
  \bibinfo {author} {\bibfnamefont {E.}~\bibnamefont {Shahmoon}}, \bibinfo
  {author} {\bibfnamefont {D.}~\bibnamefont {Oron}}, \ and\ \bibinfo {author}
  {\bibfnamefont {R.}~\bibnamefont {Ozeri}},\ }\bibfield  {title} {\enquote
  {\bibinfo {title} {Cooperative lamb shift in a mesoscopic atomic array},}\
  }\href {\doibase 10.1103/PhysRevLett.113.193002} {\bibfield  {journal}
  {\bibinfo  {journal} {Phys. Rev. Lett.}\ }\textbf {\bibinfo {volume} {113}},\
  \bibinfo {pages} {193002} (\bibinfo {year} {2014})}\BibitemShut {NoStop}%
\bibitem [{\citenamefont {Sutherland}\ and\ \citenamefont
  {Robicheaux}(2016)}]{Sut16}%
  \BibitemOpen
  \bibfield  {author} {\bibinfo {author} {\bibfnamefont {R.~T.}\ \bibnamefont
  {Sutherland}}\ and\ \bibinfo {author} {\bibfnamefont {F.}~\bibnamefont
  {Robicheaux}},\ }\bibfield  {title} {\enquote {\bibinfo {title} {Collective
  dipole-dipole interactions in an atomic array},}\ }\href {\doibase
  10.1103/PhysRevA.94.013847} {\bibfield  {journal} {\bibinfo  {journal} {Phys.
  Rev. A}\ }\textbf {\bibinfo {volume} {94}},\ \bibinfo {pages} {013847}
  (\bibinfo {year} {2016})}\BibitemShut {NoStop}%
\bibitem [{\citenamefont {Kr{\"a}mer}\ \emph {et~al.}(2016)\citenamefont
  {Kr{\"a}mer}, \citenamefont {Ostermann},\ and\ \citenamefont
  {Ritsch}}]{Kra16}%
  \BibitemOpen
  \bibfield  {author} {\bibinfo {author} {\bibfnamefont {S.}~\bibnamefont
  {Kr{\"a}mer}}, \bibinfo {author} {\bibfnamefont {L.}~\bibnamefont
  {Ostermann}}, \ and\ \bibinfo {author} {\bibfnamefont {H.}~\bibnamefont
  {Ritsch}},\ }\bibfield  {title} {\enquote {\bibinfo {title} {Optimized
  geometries for future generation optical lattice clocks},}\ }\href {\doibase
  10.1209/0295-5075/114/14003} {\bibfield  {journal} {\bibinfo  {journal}
  {{EPL} (Europhysics Letters)}\ }\textbf {\bibinfo {volume} {114}},\ \bibinfo
  {pages} {14003} (\bibinfo {year} {2016})}\BibitemShut {NoStop}%
\bibitem [{\citenamefont {do~Espirito~Santo}\ \emph {et~al.}(2020)\citenamefont
  {do~Espirito~Santo}, \citenamefont {Weiss}, \citenamefont {Cipris},
  \citenamefont {Kaiser}, \citenamefont {Guerin}, \citenamefont {Bachelard},\
  and\ \citenamefont {Schachenmayer}}]{doE20}%
  \BibitemOpen
  \bibfield  {author} {\bibinfo {author} {\bibfnamefont {T.~S.}\ \bibnamefont
  {do~Espirito~Santo}}, \bibinfo {author} {\bibfnamefont {P.}~\bibnamefont
  {Weiss}}, \bibinfo {author} {\bibfnamefont {A.}~\bibnamefont {Cipris}},
  \bibinfo {author} {\bibfnamefont {R.}~\bibnamefont {Kaiser}}, \bibinfo
  {author} {\bibfnamefont {W.}~\bibnamefont {Guerin}}, \bibinfo {author}
  {\bibfnamefont {R.}~\bibnamefont {Bachelard}}, \ and\ \bibinfo {author}
  {\bibfnamefont {J.}~\bibnamefont {Schachenmayer}},\ }\bibfield  {title}
  {\enquote {\bibinfo {title} {Collective excitation dynamics of a cold atom
  cloud},}\ }\href {\doibase 10.1103/PhysRevA.101.013617} {\bibfield  {journal}
  {\bibinfo  {journal} {Phys. Rev. A}\ }\textbf {\bibinfo {volume} {101}},\
  \bibinfo {pages} {013617} (\bibinfo {year} {2020})}\BibitemShut {NoStop}%
\bibitem [{\citenamefont {Ruostekoski}\ and\ \citenamefont
  {Javanainen}(1997)}]{Ruo97}%
  \BibitemOpen
  \bibfield  {author} {\bibinfo {author} {\bibfnamefont {Janne}\ \bibnamefont
  {Ruostekoski}}\ and\ \bibinfo {author} {\bibfnamefont {Juha}\ \bibnamefont
  {Javanainen}},\ }\bibfield  {title} {\enquote {\bibinfo {title} {Quantum
  field theory of cooperative atom response: Low light intensity},}\ }\href
  {\doibase 10.1103/PhysRevA.55.513} {\bibfield  {journal} {\bibinfo  {journal}
  {Phys. Rev. A}\ }\textbf {\bibinfo {volume} {55}},\ \bibinfo {pages}
  {513--526} (\bibinfo {year} {1997})}\BibitemShut {NoStop}%
\bibitem [{\citenamefont {Aljunid}\ \emph {et~al.}(2009)\citenamefont
  {Aljunid}, \citenamefont {Tey}, \citenamefont {Chng}, \citenamefont {Liew},
  \citenamefont {Maslennikov}, \citenamefont {Scarani},\ and\ \citenamefont
  {Kurtsiefer}}]{Alj09}%
  \BibitemOpen
  \bibfield  {author} {\bibinfo {author} {\bibfnamefont {Syed~Abdullah}\
  \bibnamefont {Aljunid}}, \bibinfo {author} {\bibfnamefont {Meng~Khoon}\
  \bibnamefont {Tey}}, \bibinfo {author} {\bibfnamefont {Brenda}\ \bibnamefont
  {Chng}}, \bibinfo {author} {\bibfnamefont {Timothy}\ \bibnamefont {Liew}},
  \bibinfo {author} {\bibfnamefont {Gleb}\ \bibnamefont {Maslennikov}},
  \bibinfo {author} {\bibfnamefont {Valerio}\ \bibnamefont {Scarani}}, \ and\
  \bibinfo {author} {\bibfnamefont {Christian}\ \bibnamefont {Kurtsiefer}},\
  }\bibfield  {title} {\enquote {\bibinfo {title} {Phase shift of a weak
  coherent beam induced by a single atom},}\ }\href {\doibase
  10.1103/PhysRevLett.103.153601} {\bibfield  {journal} {\bibinfo  {journal}
  {Phys. Rev. Lett.}\ }\textbf {\bibinfo {volume} {103}},\ \bibinfo {pages}
  {153601} (\bibinfo {year} {2009})}\BibitemShut {NoStop}%
\bibitem [{\citenamefont {Pototschnig}\ \emph {et~al.}(2011)\citenamefont
  {Pototschnig}, \citenamefont {Chassagneux}, \citenamefont {Hwang},
  \citenamefont {Zumofen}, \citenamefont {Renn},\ and\ \citenamefont
  {Sandoghdar}}]{Pot11}%
  \BibitemOpen
  \bibfield  {author} {\bibinfo {author} {\bibfnamefont {M.}~\bibnamefont
  {Pototschnig}}, \bibinfo {author} {\bibfnamefont {Y.}~\bibnamefont
  {Chassagneux}}, \bibinfo {author} {\bibfnamefont {J.}~\bibnamefont {Hwang}},
  \bibinfo {author} {\bibfnamefont {G.}~\bibnamefont {Zumofen}}, \bibinfo
  {author} {\bibfnamefont {A.}~\bibnamefont {Renn}}, \ and\ \bibinfo {author}
  {\bibfnamefont {V.}~\bibnamefont {Sandoghdar}},\ }\bibfield  {title}
  {\enquote {\bibinfo {title} {Controlling the phase of a light beam with a
  single molecule},}\ }\href {\doibase 10.1103/PhysRevLett.107.063001}
  {\bibfield  {journal} {\bibinfo  {journal} {Phys. Rev. Lett.}\ }\textbf
  {\bibinfo {volume} {107}},\ \bibinfo {pages} {063001} (\bibinfo {year}
  {2011})}\BibitemShut {NoStop}%
\bibitem [{\citenamefont {Celebrano}\ \emph {et~al.}(2011)\citenamefont
  {Celebrano}, \citenamefont {Kukura}, \citenamefont {Renn},\ and\
  \citenamefont {Sandoghdar}}]{Cel11}%
  \BibitemOpen
  \bibfield  {author} {\bibinfo {author} {\bibfnamefont {Michele}\ \bibnamefont
  {Celebrano}}, \bibinfo {author} {\bibfnamefont {Philipp}\ \bibnamefont
  {Kukura}}, \bibinfo {author} {\bibfnamefont {Alois}\ \bibnamefont {Renn}}, \
  and\ \bibinfo {author} {\bibfnamefont {Vahid.}\ \bibnamefont {Sandoghdar}},\
  }\bibfield  {title} {\enquote {\bibinfo {title} {Single-molecule imaging by
  optical absorption},}\ }\href {\doibase 10.1038/nphoton.2010.290} {\bibfield
  {journal} {\bibinfo  {journal} {Nature Photonics}\ }\textbf {\bibinfo
  {volume} {5}},\ \bibinfo {pages} {95} (\bibinfo {year} {2011})}\BibitemShut
  {NoStop}%
\bibitem [{\citenamefont {Streed}\ \emph {et~al.}(2012)\citenamefont {Streed},
  \citenamefont {Jechow}, \citenamefont {Norton},\ and\ \citenamefont
  {Kielpinski}}]{Str12}%
  \BibitemOpen
  \bibfield  {author} {\bibinfo {author} {\bibfnamefont {Erik~W.}\ \bibnamefont
  {Streed}}, \bibinfo {author} {\bibfnamefont {Andreas}\ \bibnamefont
  {Jechow}}, \bibinfo {author} {\bibfnamefont {Benjamin~G.}\ \bibnamefont
  {Norton}}, \ and\ \bibinfo {author} {\bibfnamefont {David}\ \bibnamefont
  {Kielpinski}},\ }\bibfield  {title} {\enquote {\bibinfo {title} {Absorption
  imaging of a single atom},}\ }\href {\doibase 10.1038/ncomms1944} {\bibfield
  {journal} {\bibinfo  {journal} {Nature Communications}\ }\textbf {\bibinfo
  {volume} {3}},\ \bibinfo {pages} {933} (\bibinfo {year} {2012})}\BibitemShut
  {NoStop}%
\bibitem [{\citenamefont {Alt}\ \emph {et~al.}(2003)\citenamefont {Alt},
  \citenamefont {Schrader}, \citenamefont {Kuhr}, \citenamefont {M\"uller},
  \citenamefont {Gomer},\ and\ \citenamefont {Meschede}}]{Alt03}%
  \BibitemOpen
  \bibfield  {author} {\bibinfo {author} {\bibfnamefont {Wolfgang}\
  \bibnamefont {Alt}}, \bibinfo {author} {\bibfnamefont {Dominik}\ \bibnamefont
  {Schrader}}, \bibinfo {author} {\bibfnamefont {Stefan}\ \bibnamefont {Kuhr}},
  \bibinfo {author} {\bibfnamefont {Martin}\ \bibnamefont {M\"uller}}, \bibinfo
  {author} {\bibfnamefont {Victor}\ \bibnamefont {Gomer}}, \ and\ \bibinfo
  {author} {\bibfnamefont {Dieter}\ \bibnamefont {Meschede}},\ }\bibfield
  {title} {\enquote {\bibinfo {title} {Single atoms in a standing-wave dipole
  trap},}\ }\href {\doibase 10.1103/PhysRevA.67.033403} {\bibfield  {journal}
  {\bibinfo  {journal} {Phys. Rev. A}\ }\textbf {\bibinfo {volume} {67}},\
  \bibinfo {pages} {033403} (\bibinfo {year} {2003})}\BibitemShut {NoStop}%
\bibitem [{\citenamefont {Schlosser}\ \emph {et~al.}(2001)\citenamefont
  {Schlosser}, \citenamefont {Reymond}, \citenamefont {Protsenko},\ and\
  \citenamefont {Grangier}}]{Sch01}%
  \BibitemOpen
  \bibfield  {author} {\bibinfo {author} {\bibfnamefont {Nicolas}\ \bibnamefont
  {Schlosser}}, \bibinfo {author} {\bibfnamefont {Georges}\ \bibnamefont
  {Reymond}}, \bibinfo {author} {\bibfnamefont {Igor}\ \bibnamefont
  {Protsenko}}, \ and\ \bibinfo {author} {\bibfnamefont {Philippe}\
  \bibnamefont {Grangier}},\ }\bibfield  {title} {\enquote {\bibinfo {title}
  {Sub-poissonian loading of single atoms in a microscopic dipole trap},}\
  }\href {\doibase 10.1038/35082512} {\bibfield  {journal} {\bibinfo  {journal}
  {Nature}\ }\textbf {\bibinfo {volume} {411}},\ \bibinfo {pages} {1024--1027}
  (\bibinfo {year} {2001})}\BibitemShut {NoStop}%
\bibitem [{\citenamefont {Bruno}\ \emph {et~al.}(2019)\citenamefont {Bruno},
  \citenamefont {Bianchet}, \citenamefont {Prakash}, \citenamefont {Li},
  \citenamefont {Alves},\ and\ \citenamefont {Mitchell}}]{Bru19}%
  \BibitemOpen
  \bibfield  {author} {\bibinfo {author} {\bibfnamefont {Natalia}\ \bibnamefont
  {Bruno}}, \bibinfo {author} {\bibfnamefont {Lorena~C.}\ \bibnamefont
  {Bianchet}}, \bibinfo {author} {\bibfnamefont {Vindhiya}\ \bibnamefont
  {Prakash}}, \bibinfo {author} {\bibfnamefont {Nan}\ \bibnamefont {Li}},
  \bibinfo {author} {\bibfnamefont {Nat\'{a}lia}\ \bibnamefont {Alves}}, \ and\
  \bibinfo {author} {\bibfnamefont {Morgan~W.}\ \bibnamefont {Mitchell}},\
  }\bibfield  {title} {\enquote {\bibinfo {title} {Maltese cross coupling to
  individual cold atoms in free space},}\ }\href {\doibase
  10.1364/OE.27.031042} {\bibfield  {journal} {\bibinfo  {journal} {Opt.
  Express}\ }\textbf {\bibinfo {volume} {27}},\ \bibinfo {pages} {31042--31052}
  (\bibinfo {year} {2019})}\BibitemShut {NoStop}%
\bibitem [{\citenamefont {Fernandes}\ \emph {et~al.}(2012)\citenamefont
  {Fernandes}, \citenamefont {Sievers}, \citenamefont {Kretzschmar},
  \citenamefont {Wu}, \citenamefont {Salomon},\ and\ \citenamefont
  {Chevy}}]{Fer12}%
  \BibitemOpen
  \bibfield  {author} {\bibinfo {author} {\bibfnamefont {D.~Rio}\ \bibnamefont
  {Fernandes}}, \bibinfo {author} {\bibfnamefont {F.}~\bibnamefont {Sievers}},
  \bibinfo {author} {\bibfnamefont {N.}~\bibnamefont {Kretzschmar}}, \bibinfo
  {author} {\bibfnamefont {S.}~\bibnamefont {Wu}}, \bibinfo {author}
  {\bibfnamefont {C.}~\bibnamefont {Salomon}}, \ and\ \bibinfo {author}
  {\bibfnamefont {F.}~\bibnamefont {Chevy}},\ }\bibfield  {title} {\enquote
  {\bibinfo {title} {{Sub-Doppler laser cooling of fermionic
  \textsuperscript{40}K atoms in three-dimensional gray optical molasses}},}\
  }\href {\doibase 10.1209/0295-5075/100/63001} {\bibfield  {journal} {\bibinfo
   {journal} {{EPL} (Europhysics Letters)}\ }\textbf {\bibinfo {volume}
  {100}},\ \bibinfo {pages} {63001} (\bibinfo {year} {2012})}\BibitemShut
  {NoStop}%
\bibitem [{\citenamefont {Grier}\ \emph {et~al.}(2013)\citenamefont {Grier},
  \citenamefont {Ferrier-Barbut}, \citenamefont {Rem}, \citenamefont
  {Delehaye}, \citenamefont {Khaykovich}, \citenamefont {Chevy},\ and\
  \citenamefont {Salomon}}]{Gri13}%
  \BibitemOpen
  \bibfield  {author} {\bibinfo {author} {\bibfnamefont {Andrew~T.}\
  \bibnamefont {Grier}}, \bibinfo {author} {\bibfnamefont {Igor}\ \bibnamefont
  {Ferrier-Barbut}}, \bibinfo {author} {\bibfnamefont {Benno~S.}\ \bibnamefont
  {Rem}}, \bibinfo {author} {\bibfnamefont {Marion}\ \bibnamefont {Delehaye}},
  \bibinfo {author} {\bibfnamefont {Lev}\ \bibnamefont {Khaykovich}}, \bibinfo
  {author} {\bibfnamefont {Fr\'ed\'eric}\ \bibnamefont {Chevy}}, \ and\
  \bibinfo {author} {\bibfnamefont {Christophe}\ \bibnamefont {Salomon}},\
  }\bibfield  {title} {\enquote {\bibinfo {title}
  {$\ensuremath{\Lambda}$-enhanced sub-doppler cooling of lithium atoms in
  ${D}_{1}$ gray molasses},}\ }\href {\doibase 10.1103/PhysRevA.87.063411}
  {\bibfield  {journal} {\bibinfo  {journal} {Phys. Rev. A}\ }\textbf {\bibinfo
  {volume} {87}},\ \bibinfo {pages} {063411} (\bibinfo {year}
  {2013})}\BibitemShut {NoStop}%
\bibitem [{\citenamefont {Brown}\ \emph {et~al.}(2019)\citenamefont {Brown},
  \citenamefont {Thiele}, \citenamefont {Kiehl}, \citenamefont {Hsu},\ and\
  \citenamefont {Regal}}]{Bro19}%
  \BibitemOpen
  \bibfield  {author} {\bibinfo {author} {\bibfnamefont {M.~O.}\ \bibnamefont
  {Brown}}, \bibinfo {author} {\bibfnamefont {T.}~\bibnamefont {Thiele}},
  \bibinfo {author} {\bibfnamefont {C.}~\bibnamefont {Kiehl}}, \bibinfo
  {author} {\bibfnamefont {T.-W.}\ \bibnamefont {Hsu}}, \ and\ \bibinfo
  {author} {\bibfnamefont {C.~A.}\ \bibnamefont {Regal}},\ }\bibfield  {title}
  {\enquote {\bibinfo {title} {Gray-molasses optical-tweezer loading:
  Controlling collisions for scaling atom-array assembly},}\ }\href {\doibase
  10.1103/PhysRevX.9.011057} {\bibfield  {journal} {\bibinfo  {journal} {Phys.
  Rev. X}\ }\textbf {\bibinfo {volume} {9}},\ \bibinfo {pages} {011057}
  (\bibinfo {year} {2019})}\BibitemShut {NoStop}%
\bibitem [{Note1()}]{Note1}%
  \BibitemOpen
  \bibinfo {note} {We have checked numerically that the shifts calculated by
  the steady-state model is the same as the one obtained by the time-dependent
  coupled-dipole model}\BibitemShut {NoStop}%
\bibitem [{Sup()}]{SupMat}%
  \BibitemOpen
  \href@noop {} {\enquote {\bibinfo {title} {Supplemental material},}\
  }\BibitemShut {NoStop}%
\bibitem [{Note2()}]{Note2}%
  \BibitemOpen
  \bibinfo {note} {In 1D, the field of the dipoles scattered along the chain
  are also $\sigma _+$-polarized, and therefore one does not need to resort to
  the application of a large magnetic field to isolate a two-level structure,
  as was done for a random ensemble~\cite {Jen18}. In this case the Green's
  function $G({\protect \bf r})$ is also a scalar~\cite {SupMat}.}\BibitemShut
  {Stop}%
\bibitem [{\citenamefont {Friedberg}\ \emph {et~al.}(1973)\citenamefont
  {Friedberg}, \citenamefont {Hartmann},\ and\ \citenamefont
  {Manassah}}]{Fri73}%
  \BibitemOpen
  \bibfield  {author} {\bibinfo {author} {\bibfnamefont {R.}~\bibnamefont
  {Friedberg}}, \bibinfo {author} {\bibfnamefont {S.R.}\ \bibnamefont
  {Hartmann}}, \ and\ \bibinfo {author} {\bibfnamefont {J.T.}\ \bibnamefont
  {Manassah}},\ }\bibfield  {title} {\enquote {\bibinfo {title} {Frequency
  shifts in emission and absorption by resonant systems ot two-level atoms},}\
  }\href {\doibase https://doi.org/10.1016/0370-1573(73)90001-X} {\bibfield
  {journal} {\bibinfo  {journal} {Physics Reports}\ }\textbf {\bibinfo {volume}
  {7}},\ \bibinfo {pages} {101 -- 179} (\bibinfo {year} {1973})}\BibitemShut
  {NoStop}%
\bibitem [{Note3()}]{Note3}%
  \BibitemOpen
  \bibinfo {note} {Note that although it may look like this argument implies
  guiding of the light along the chain, guided modes exist only for a spacing
  smaller than $\lambda _0/2$ see \cite {Bet16a,Ase17}.}\BibitemShut {Stop}%
\bibitem [{\citenamefont {Grynberg}\ \emph {et~al.}(2010)\citenamefont
  {Grynberg}, \citenamefont {Aspect},\ and\ \citenamefont {Fabre}}]{GAF}%
  \BibitemOpen
  \bibfield  {author} {\bibinfo {author} {\bibfnamefont {Gilbert}\ \bibnamefont
  {Grynberg}}, \bibinfo {author} {\bibfnamefont {Alain}\ \bibnamefont
  {Aspect}}, \ and\ \bibinfo {author} {\bibfnamefont {Claude}\ \bibnamefont
  {Fabre}},\ }\href@noop {} {\emph {\bibinfo {title} {Introduction to Quantum
  Optics}}}\ (\bibinfo  {publisher} {Cambridge University Press},\ \bibinfo
  {year} {2010})\BibitemShut {NoStop}%
\bibitem [{\citenamefont {Kr{\"a}mer}\ and\ \citenamefont
  {Ritsch}(2015)}]{Kra15}%
  \BibitemOpen
  \bibfield  {author} {\bibinfo {author} {\bibfnamefont {Sebastian}\
  \bibnamefont {Kr{\"a}mer}}\ and\ \bibinfo {author} {\bibfnamefont {Helmut}\
  \bibnamefont {Ritsch}},\ }\bibfield  {title} {\enquote {\bibinfo {title}
  {Generalized mean-field approach to simulate the dynamics of large open spin
  ensembles with long range interactions},}\ }\href {\doibase
  10.1140/epjd/e2015-60266-5} {\bibfield  {journal} {\bibinfo  {journal} {The
  European Physical Journal D}\ }\textbf {\bibinfo {volume} {69}},\ \bibinfo
  {pages} {282} (\bibinfo {year} {2015})}\BibitemShut {NoStop}%
\bibitem [{\citenamefont {Parmee}\ and\ \citenamefont {Cooper}(2018)}]{Par18}%
  \BibitemOpen
  \bibfield  {author} {\bibinfo {author} {\bibfnamefont {C.~D.}\ \bibnamefont
  {Parmee}}\ and\ \bibinfo {author} {\bibfnamefont {N.~R.}\ \bibnamefont
  {Cooper}},\ }\bibfield  {title} {\enquote {\bibinfo {title} {Phases of driven
  two-level systems with nonlocal dissipation},}\ }\href {\doibase
  10.1103/PhysRevA.97.053616} {\bibfield  {journal} {\bibinfo  {journal} {Phys.
  Rev. A}\ }\textbf {\bibinfo {volume} {97}},\ \bibinfo {pages} {053616}
  (\bibinfo {year} {2018})}\BibitemShut {NoStop}%
\bibitem [{\citenamefont {Lehmberg}(1970)}]{Leh70}%
  \BibitemOpen
  \bibfield  {author} {\bibinfo {author} {\bibfnamefont {R.~H.}\ \bibnamefont
  {Lehmberg}},\ }\bibfield  {title} {\enquote {\bibinfo {title} {{Radiation
  from an N-atom system. I. General formalism}},}\ }\href {\doibase
  10.1103/PhysRevA.2.883} {\bibfield  {journal} {\bibinfo  {journal} {Physical
  Review A}\ }\textbf {\bibinfo {volume} {2}},\ \bibinfo {pages} {883--888}
  (\bibinfo {year} {1970})}\BibitemShut {NoStop}%
\bibitem [{\citenamefont {Gross}\ and\ \citenamefont {Haroche}(1982)}]{Gro82}%
  \BibitemOpen
  \bibfield  {author} {\bibinfo {author} {\bibfnamefont {M.}~\bibnamefont
  {Gross}}\ and\ \bibinfo {author} {\bibfnamefont {S.}~\bibnamefont
  {Haroche}},\ }\bibfield  {title} {\enquote {\bibinfo {title} {Superradiance:
  An essay on the theory of collective spontaneous emission},}\ }\href
  {\doibase https://doi.org/10.1016/0370-1573(82)90102-8} {\bibfield  {journal}
  {\bibinfo  {journal} {Physics Reports}\ }\textbf {\bibinfo {volume} {93}},\
  \bibinfo {pages} {301 -- 396} (\bibinfo {year} {1982})}\BibitemShut {NoStop}%
\end{thebibliography}%

	\clearpage

\begin{widetext}
\begin{center}
	{\Large Supplemental material}
\end{center} 
\appendix

\section{Derivation of eq.~(1) of the main text}

As said in the main text, in the weak excitation limit, an ensemble of driven two-level atoms can be treated as coupled damped harmonic oscillators. In steady state, this leads to a set of linear coupled equations:
\begin{equation}
0 = (2i\delta-1) D_n -E_L(\boldsymbol{r}_n) - \sum_{m \neq n}g(\boldsymbol{r}_m-\boldsymbol{r}_n) D_m\ ,
\label{steady}
\end{equation}
where we have defined $D_n=d_n\,i\,(k_0^3/6\pi\varepsilon_0)$. Here $\delta = \Delta/\Gamma_0$ is the detuning normalized by the natural linewidth  $\Gamma_0$, and $g(\boldsymbol{r})=G(\boldsymbol r)\,i(6\pi\varepsilon_0/k_0^3)$ is the dimensionless field propagator:
\begin{equation}
g(\boldsymbol{r}) = \frac{3e^{ikr}}{2i} \left\lbrace [3|\hat{\boldsymbol{r}}.\hat{\boldsymbol{e}}_y |^2-1 ]\left[ \frac{1}
{(kr)^3}-\frac{i}{(kr)^2} \right] + [1-|\hat{\boldsymbol{r}}.\hat{\boldsymbol{e}}_y |^2]\frac{1}{kr} \right\rbrace. 
\label{propagator}
\end{equation}
In our system, with $\hat{\boldsymbol{r}}\cdot\hat{\boldsymbol{e}}_y = 0$, the expression of the field propagator 
\eqref{propagator} simplifies to

\begin{equation}
g(\boldsymbol{r}) = \frac{3e^{ikr}}{2i} \left[ \frac{1}{kr}+\frac{i}{(kr)^2}-\frac{1}{(kr)^3} \right]. \label{propagator1D}
\end{equation}
Equation \eqref{steady} is then
\begin{equation}
(2i\delta-1)D_n - \sum_{m\neq n} \frac{3e^{ik|z_m-z_n|}}{2ik|z_m-z_n|}f_{mn}D_m = E_{\rm L}e^{ikz_n}. \label{coupled}
\end{equation}
with $f_{mn}=1+\frac{i}{k|z_m-z_n|}-\frac{1}{(k|z_m-z_n|)^2}$. Assuming $k|z_m-z_n| \gg 1$ for all $m\neq n$, one can solve perturbatively the equations \eqref{coupled} and get the zeroth order dipoles
\begin{equation}
D_n^{(0)} = \frac{E_{\rm L}e^{ikz_n}}{2i\delta-1}.
\end{equation}
The field at first order is then 
\begin{equation}
	E^{(1)}(z)= E_{\rm L}e^{ikz}+ \sum_{m\neq n} \frac{3e^{ik|z_m-z|}}{2ik|z_m-z|}\frac{E_{\rm L}e^{ikz_m}}{2i\delta-1}
\end{equation}
%
Factorizing the global phase $e^{ikz}$ and splitting the sum into two parts leads to

\begin{equation}
E^{(1)}(z)= E_{\rm L}e^{ikz}\left\lbrace 1+ \frac{3}{2i(2i\delta-1)}\left[ \underbrace{\sum_{z_m<z}
\frac{1}{k(z-z_m)}}_{\rm{forward\:scattering}} + \underbrace{\sum_{z_m>z}\frac{e^{2ik(z_m-
z)}}{k(z_m-z)}}_{\rm{backward\:scattering}} \right] \right\rbrace \label{firstOrder}
\end{equation}
Equation \eqref{firstOrder} shows that all forward scattering terms add constructively at long distance. The backward scattering is suppressed at long range because of the random phase factor $2k (z_m-z)$. Keeping only the forward scattering term, the field intensity at position $z$ is thus given by
\begin{equation}
|E^{(1)}(z)|^2 = |E_{\rm L}|^2\left| 1+ \frac{3}{2i(2i\delta-1)}\sum_{z_m<z}\frac{1}{k(z-z_m)} \right|^2
\end{equation}
which gives at lowest order in $1/k(z-z_m)$,
\begin{equation}
|E^{(1)}(z)|^2 = |E_{\rm L}|^2\left[ 1- \frac{6\delta}{1+4\delta^2}\sum_{z_m<z}\frac{1}{k(z-z_m)} \right].
\end{equation}

\section{Derivation of nonlinear coupled dipole model (NCD)}
Consider an ensemble of $N$ atoms located at $\mathbf{r}_i$ in space. Each atom has two levels $|\rm e\rangle$ and $|\rm g\rangle$ differing in energy by $\hbar\omega_{\rm e}$, coupled via a $\sigma^+$ transition with dipole moment $d$. The atoms are coupled via electromagnetic field modes enumerated by $q$. If we label the mode occupation number operator as $a_q^\dagger a_q$, we can write the system Hamiltonian as ($\hbar = 1$)
\begin{equation}
 \mathcal{H} = \sum_{n=1}^N \omega_{\rm e} |{\rm e}_n\rangle\langle {\rm e}_n| + \sum_q \omega_q a_q^\dagger a_q
 -\sum_{n=1}^N \sum_q \sqrt{\frac{2\pi \omega_q}{V}}  \left(\mathbf{\hat{\varepsilon}}_q \mathrm{e}^{i \mathbf{k}_q\cdot \mathbf{r}_n} a_q
 +\mathbf{\hat{\varepsilon}}_q^* \mathrm{e}^{-i \mathbf{k}_q \mathbf{r}_n}a_q^\dagger \right)
(\mathbf{d}_{+,n}+\mathbf{d}_{-,n}),
\end{equation}
where the dipole operator of the $n$-th atom can be written as
\begin{equation}
\mathbf{d}_n = \underbrace{\frac{d}{\sqrt{2}}\begin{pmatrix}
1\\ 
i\\
 0
 \end{pmatrix}|{\rm e}_n\rangle\langle {\rm g}_n|}
_{\equiv \mathbf{d}_{+}}
+\underbrace{\frac{d}{\sqrt{2}}\begin{pmatrix}
1\\ 
-i\\
 0
 \end{pmatrix}|{\rm g}_n\rangle\langle {\rm e}_n|
}_{\equiv \mathbf{d}_{-}}.
\end{equation}

Following the approach of R.~H.~Lehmberg~\cite{Leh70}, we integrate out the internal photon hopping between dipoles via EM modes, and obtain~\cite{Jon16, Ase17} in the rotating-wave approximation the evolution of the $N$-atom density matrix $\rho$
\begin{equation}
\frac{\mathrm{d}\rho}{\mathrm{d}t} = i[\mathcal{H}_{\rm eff},\rho]+\mathcal{L}[\rho],
\end{equation}
where
\begin{eqnarray}\label{eq:quantum}
\mathcal{H}_{\rm eff} &=& -\sum_{n=1}^N \Delta|{\rm e}_n \rangle\langle {\rm e}_n|
+ \sum_{n=1}^N  \left(\mathbf{d}_{+,n}\cdot\mathbf{E}^{(0)}_+
+\mathbf{d}_{-,n}\cdot\mathbf{E}^{(0)\dagger}_+\right)
-\sum_{i,j=x,y,z}~\sum_{n,m=1}^N d^i_{+,n}\cdot \overset\leftrightarrow{V}^{ij}_{nm}\cdot d^j_{-,m},\\
\mathcal{L}[\rho] &=& \sum_{i,j}\sum_{n,m}\overset\leftrightarrow{\Gamma}^{ij}_{nm}\left[ d^j_{-,m}~ \rho~d^i_{+,n}
-\frac{1}{2} \left\{d^i_{+,n}~d^j_{-,m},~\rho \right\} \right].
\end{eqnarray}   
The effective dipole-dipole interaction $\overset\leftrightarrow{V}_{nm} = \mathrm{Re}\left[ \overset\leftrightarrow{G}(\mathbf{r}_{nm})\right]$
and decay $\overset\leftrightarrow{\Gamma}_{nm} = 2 ~\mathrm{Im}\left[  \overset\leftrightarrow{G}(\mathbf{r}_{nm}) \right],
$ (with $\mathbf{r}_{nm}=\mathbf{r}_m-\mathbf{r}_n$) depend on the Green's dyadic function
\begin{equation}
\overset\leftrightarrow{G}(\mathbf{r})=  \frac{k_0^3}{4\pi\varepsilon_0} \mathrm{e}^{ikr}
\left\{ \left[ \frac{1}{kr}+\frac{i}{(kr)^2}-\frac{1}{(kr)^3} \right] \mathbb{•}{1} 
+\left[ -\frac{1}{kr}-\frac{3i}{(kr)^2} +\frac{3}{(kr)^3}\right]  \left|\hat{\mathbf{r}} \rangle\langle\hat{\mathbf{r}}\right|
\right\}.
\end{equation}
We will now assume that
\emph{quantum} correlations that form between the atoms
are negligible and solve the above set of equations in the mean-field limit
$\rho = \bigotimes_n \rho_n$, where $\rho_n$ are density matrices of individual atoms. We can then solve the above system for the coherence
$\rho_{{\rm eg},n}$ and excited state population $\rho_{{\rm ee}, n}$ of atom $n$
\begin{eqnarray}
\frac{\mathrm{d}\rho_{{\rm ee},n}}{\mathrm{d}t} &=&
 -\Gamma_0\rho_{{\rm ee},n} + i\rho_{{\rm ge},n} \frac{\Omega_n}{2} -i \rho_{{\rm eg},n} \frac{\Omega_n^*}{2} \label{eq:nqd1} \\
 \frac{\mathrm{d}\rho_{{\rm eg},n}}{\mathrm{d}t} &=&  
 i\Delta \rho_{{\rm eg  },n} - \frac{\Gamma_0}{2} \rho_{{\rm eg},n}
 - i(\rho_{{\rm ee},n}-\rho_{{\rm gg},n})\frac{\Omega_n}{2} \label{eq:nqd2}
\end{eqnarray}
where the driving Rabi frequency for dipole $n$ is given by the sum of the laser driving field and field scattered by other dipoles
\begin{equation}
\frac{\Omega_n}{2} = \mathbf{d}_{+,n} \cdot
\left[ \mathbf{E}_+(\mathbf{r}_n) + \sum_{m\neq n} \overset\leftrightarrow{G}(\mathbf{r}_n-\mathbf{r}_m) \mathbf{d}_{-,m}\right].
\end{equation}
The solution of Eqs.~(\ref{eq:nqd1}-\ref{eq:nqd2}) has the same form as
that of a single driven atom with Rabi frequency $\Omega_n$
\begin{equation}
\rho_{{\rm eq},n} = \frac{i \Omega_n}{\Gamma_0}x
\frac{1+2i\Delta/\Gamma_0}{1+4(\Delta/\Gamma_0)^2 + 2\Omega_n^2/\Gamma_0^2}.
\end{equation}

By keeping the full evolution of a single quantum emitter we
account for the saturation of the emitted field from the individual emitters under strong driving. Considering the large interatomic spacing in the experimental conditions we do not expect
strong correlations to occur due to the driving field. We do however
neglect correlations that can occur due to collective decay:
Cascades of such collective decays can lead to a well
defined phase between spatially distant atoms, giving rise to superradiant phenomena for example \cite{Gro82}. However collective decays can be neglected because under persistent laser driving there
is not enough time for a cascade to build up strong enough coherence
between individual atoms. If the driving was pulsed, the dynamics might experience strong quantum correlations. It is not the case here.

To further check the validity of our approximation we compare the predictions
of the NCD model with the full quantum evolution given by Eq.~(\ref{eq:quantum})
integrated using quantum Monte Carlo as in Ref.~\cite{Jon16,Ase17}. Results for unity-filled chains with the inter-site distance $a=\SI{470}{\nano\meter}$ as in the experiment are shown in Fig.\,4(a) of the main text. They are in relatively good agreement with NCD predictions. However, we do observe differences showing that correlations (correctly accounted for by quantum Monte Carlo) could be playing a role, even at these relatively large distances. This will be the topic of future work. 

\section{Reduction of the relative line shift under strong driving}

For $I/I_{\rm sat} \gtrsim 1$, the linear coupled-dipole model fails because
its prediction for the atomic dipole [Fig.~\ref{fig:sup}(a) dashed line]
does not capture the saturation of two-level
quantum emitters~[Fig.~\ref{fig:sup}(a) solid line].
The linear model predicts how large the atomic dipoles should be if the
shift of the fluorescence peak $\delta \omega$ was to remain the same
even for increased driving amplitudes $\Omega$.
The dipole saturation is the main cause for the reduction of the shift. The shift relative to its value for vanishing drive $\delta \omega(\Omega)/\delta \omega(\Omega\rightarrow 0)$ has the same behaviour for all atom numbers $N$ for unity-filled chains in the NCD model as can be seen Fig.~\ref{fig:sup}(b) where curves for different atom numbers collapse in a single one.\\
\begin{figure}[htbp]
\includegraphics[width=.6\linewidth]{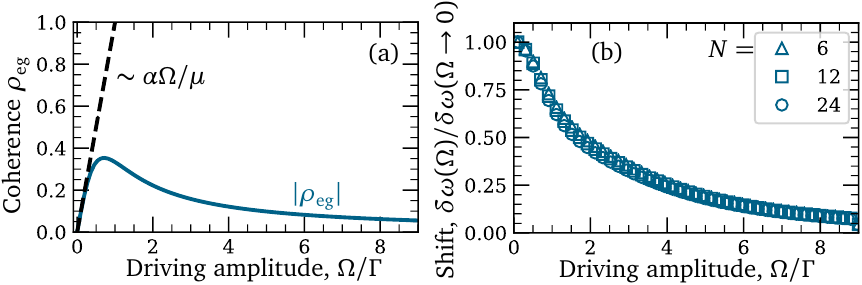} 
\caption{\label{fig:sup}
For strong driving, the coherence $\rho_{\rm eg}$ of a two-level quantum 
system saturates (a, solid line), departing from the classical linear-dipole
 prediction (a, dashed line). The mean-field
 non-linear coupled-dipole model (NCD) calculations
 for different unity-filled $N$-atom chains~(b) predicts that the relative 
 shift will be reduced in the same manner for different chain lengths.}
\end{figure}

\section{Effect of random filling and position disorder in each well}
For a perfectly filled lattice with no disorder, one expects two things: First, an overall gradual shift due to a variation in the strength of the interaction akin to what we observe when changing the filling in the main text. Second a modulation due to a Fabry-Perot effect when the lattice constant is a multiple of half the wavelength (the phase shift between sites is then a multiple of $\pi$), this is discussed for instance in \cite{Sut16}. In the case of a randomly filled and disordered chain, the shift is reduced by the radial disorder and the modulations wash out because the Fabry-Perot effect is blurred by the axial disorder, but some modulation remains. This interesting effect could be the future of future research. We compare the predictions for a randomly filled, disordered chain to that of a perfect chain with an atom placed every other site in Fig.\,\ref{fig:sup1}. For our lattice constant of 1.2 wavelengths, the Fabry-Perot effect is absent and there is no strong effect of the disorder except a small reduction of the shift.
\begin{figure}[htbp]
\includegraphics[width=0.4\linewidth]{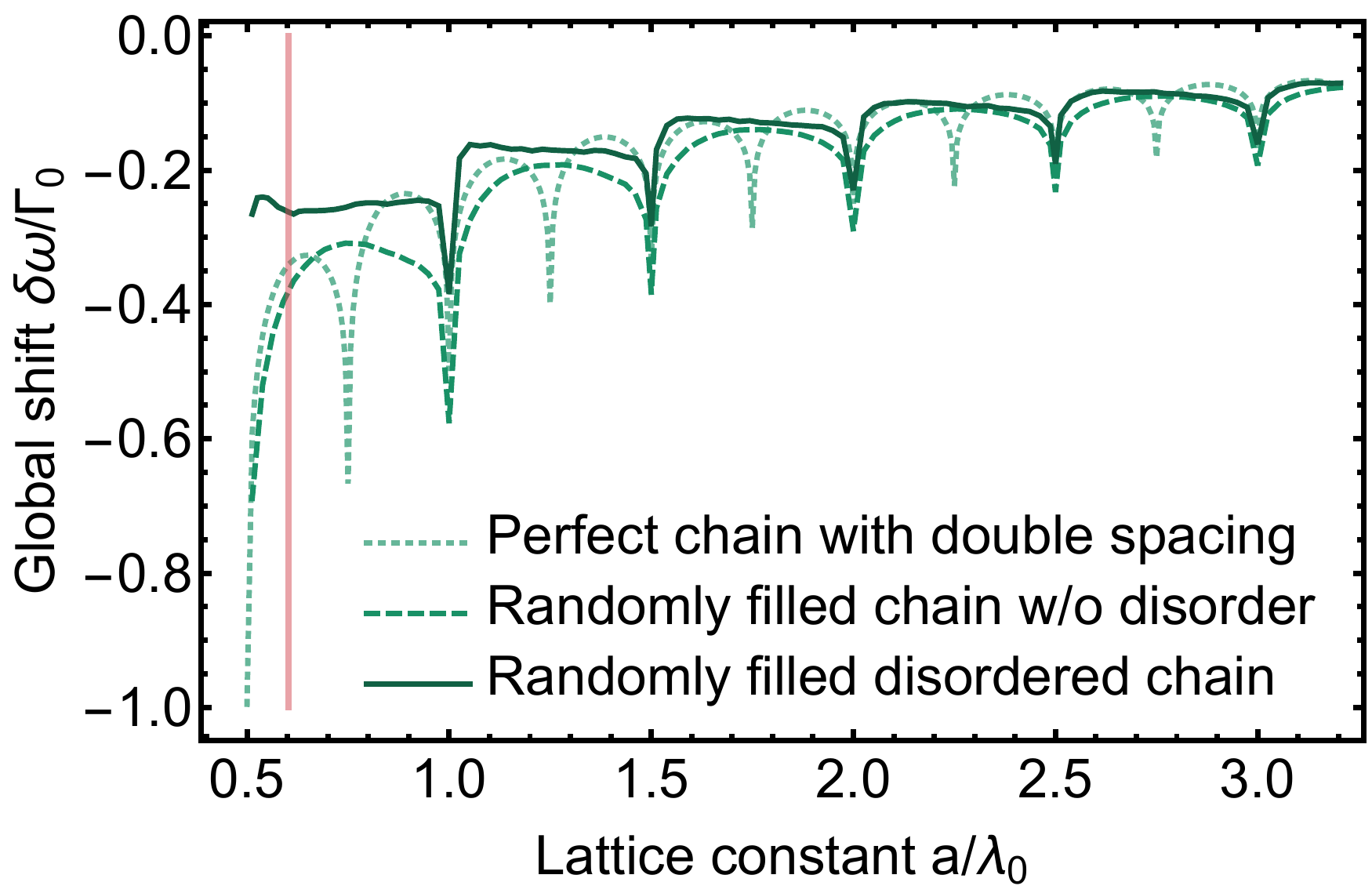}
\caption{Predicted global shift $\delta \omega/\,\Gamma_0$ as a function of distance between two lattice sites for different situations: randomly filled chain with position disorder in each well (as in the experiments), randomly filled chain without disorder and perfect chain (with double spacing to keep the same average nearest-neighbour distance). We indicate as a vertical line the spacing ($a/\lambda_0=470/780\simeq0.6$) of the experiments presented in this paper.  \label{fig:sup1}
}
\end{figure}

\section{Methods for the measurement of the scattered light}

We excite the  cloud  with 50\,pulses of 200\,ns in free space, i.e.~with the dipole trap switched off. The probe has the same polarization ($\sigma_+$) as the optical pumping beam (it is recombined before an optical fiber and the polarization is filtered afterwards with a polarizing beam-splitter cube), the magnetic field is kept the same during the optical pumping as during the probing. Between the probe pulses, we recapture the atoms by applying 200\,ns pulses of the trapping laser. This method ensures that the chain does not expand significantly during the probing time and retains its 1D symmetry. The release-and-recapture rate is high enough to avoid the parametric heating of the atoms. The number of probe pulses and their duration is set such that the signal-to-noise ratio is acceptable, without resulting into heating and atom losses higher than 5\,\%. We have checked the heating and losses for the maximum probe intensity used in the paper ($\Omega/\Gamma_0\simeq3$, corresponding to about $I=20\,I_{\rm sat}$), and used this pulse number for all intensities (lower intensities result in weaker heating).

The pulse cycles induce a change in effective trapping. We have verified by running simulations of the atomic motion under the pulse cycles that the increase in the transverse width of the cloud leads to an expected reduction of the measured global collective shift of about 6\,\%, which is much less than our experimental uncertainties.

 For the weak excitation regime, we set the probe beam intensity to $\frac{I}{I_{sat}}\simeq 0.3$, while for the data of Fig.~4, the probe intensity is varied. The transverse scattered light is collected on an electron-multiplying CCD. It is then recorded as a function of the frequency of the excitation beam in a range $[-3\Gamma_0,3\Gamma_0]$ around the single atom resonance frequency where $\Gamma_0 = 2\pi\times\SI{6}{\mega\hertz}$ is the natural linewidth. The sequence is repeated 300 times. The resonance frequency is then extracted using a lorentzian fit over the data.

\section{Measurement of the radial size of the cloud}

The transverse imaging axis gives access directly to the radial and axial sizes of the cloud for sizes larger than about $\SI{1}{\micro\meter}$ using a 2D gaussian fit of the integrated fluorescence.  We let the cloud expand in free flight during $t_{\rm{of}}$ and measure its size. The radial size is $ \sigma_{\rho}(t_{\rm{of}}) = \sqrt{\sigma_{\rho,0}^2 + v_{T}^2 t_{\rm{of}}^2}$ where 
$v_{T} = \sqrt{\frac{k_{\rm{B}}T}{m}}$  is the atomic velocity. By repeating the measurement for different $t_{\rm{of}}$, we obtain the temperature of the atoms: T$\,=\SI{80\pm20}{\micro\kelvin}$. Using the trap parameters, this gives us the initial radial size $\sigma_{\rho,0} = \SI{285 \pm 30}{\nano\meter}$ and thus the radial size of the cloud at all times.

\section{Control of the average interatomic distance in the chain}

The interatomic distance is changed in the following way: a full chain is loaded as explained in the main text. Some fraction of the atoms is first optically pumped in the $(5^2S_{1/2},\,F=2)$ state. These pumped atoms are then ejected from the trap using a state selective ``push-out beam'', that is resonant with the $(5^2S_{1/2},\,F=2)\rightarrow(5^2P_{3/2},\,F'=3)$. The remaining atoms are assumed to form a uniformly filled chain, whose length is equal to the initial one but the density is smaller. To measure the filling, we assume that for a saturating probe intensity, the fluorescence is proportional to the filling. The relation between fluorescence and filling is obtained independently using the calibration of the fluorescence of a single atom. The standard deviation of the filling is estimated by collecting the fluorescence of a 5ms pulse of the MOT beams just after the loading of the chain. This method is prone to systematic errors due to the light shift of the trapping laser, but it allows us to estimate the relative shot-to-shot fluctuations of the filling. On the plot Fig.\,3(b), the reference of the shifts is defined as the intercept of a linear fit of the data, and the value is compatible with the non-interacting reference value of Fig.\,3(a). The average interatomic distance between closest neighbour atoms for the time-of-flight experiment [red squares in the insert Fig.\,3(b)] is simulated by choosing randomly the positions of the atoms with a uniform distribution along the chain axis and according to a Gaussian law with variance $\sigma_{\rho}$ in the radial direction.

\section{Reference of the measured shifts}
The frequency of lasers is locked by standard saturated absorption spectroscopy in Rb cells. However, laser frequency drifts, for instance in lock electronics, RF sources for frequency shifters etc. might lead to day-to-day drifts in the experiments. To prevent from such systematics, we verify the position of the resonance of independent \emph{non-interacting} atoms. The reference we use for this is that of 3D dilute clouds after a long time-of-flight such that interactions do not play a role. We have verified that in this case, we do not find a shift between the resonance when exciting using the probe along the cloud axis ($\hat{\boldsymbol{z}}$) and perpendicularly to it, as opposed to what is found in the trapped case. This allows us to define the shift in the different figures of the paper. In the data of Fig.\,3(b), no reference could be taken on the same day data was taken, the reference is defined as the intercept of a linear fit to the data with the $\eta=0$ axis, the reference value this obtained corresponds to the reference value obtained by spectroscopy of noninteracting atoms on other days. 
\end{widetext}
\end{document}